\documentclass[aip,jcp,reprint,floatfix,nofootinbib]{revtex4-2}

\usepackage[utf8x]{inputenc}
\usepackage{mathtools}
\usepackage{fixltx2e}
\usepackage[english]{babel}

\usepackage{letltxmacro}
\LetLtxMacro{\ORIGselectlanguage}{\selectlanguage}
\makeatletter
\DeclareRobustCommand{\selectlanguage}[1]{%
  \@ifundefined{alias@\string#1}
    {\ORIGselectlanguage{#1}}
    {\begingroup\edef\x{\endgroup
       \noexpand\ORIGselectlanguage{\@nameuse{alias@#1}}}\x}%
}
\newcommand{\definelanguagealias}[2]{%
  \@namedef{alias@#1}{#2}%
}
\makeatother

\definelanguagealias{en}{english}

\usepackage{graphicx}
\graphicspath{{figures/}}
\usepackage{physics}
\usepackage{multirow}
\usepackage{xcolor}
\usepackage{caption}
\usepackage{subcaption}
\usepackage{xcolor}
\usepackage{bm}
\usepackage{yfonts} 
\usepackage[title,titletoc]{appendix}
\usepackage{diagbox}
\usepackage[algoruled]{algorithm2e}
\usepackage{manfnt}
\usepackage{tensor} 
\usepackage{dcolumn}  
\usepackage{verbatim}
\usepackage{suffix}
\usepackage{booktabs}
\usepackage{tikz}
\usepackage{tikz-network}
\usetikzlibrary{tikzmark}
\newlength{\mytextsize}
\newcommand{\JoinUp}[5]{\begin{tikzpicture}[remember picture,overlay,line width=0.05\mytextsize]
    \draw([shift={(#1\mytextsize,#2\mytextsize)}]pic cs:start#5) -- ++(0pt,0.7\mytextsize) -| ([shift={(#3\mytextsize,#4\mytextsize)}]pic cs:end#5);
    \end{tikzpicture}}

\newcommand{\JoinUpDown}[5]{\begin{tikzpicture}[remember picture,overlay,line width=0.05\mytextsize]
    \draw([shift={(#1\mytextsize,#2\mytextsize)}]pic cs:start#5) -- ++(0pt,0.7\mytextsize) -| (pic cs:middle#5) -| ++(0pt,-1.5\mytextsize) -| ([shift={(#3\mytextsize,-#4\mytextsize)}]pic cs:end#5);
    \end{tikzpicture}}
\newcommand{\JoinDownUp}[5]{\begin{tikzpicture}[remember picture,overlay,line width=0.05\mytextsize]
    \draw([shift={(#1\mytextsize,#2\mytextsize)}]pic cs:start#5) -- ++(0pt,-0.7\mytextsize) -| (pic cs:middle#5) -| ++(0pt,1.4\mytextsize - #2\mytextsize) -| ([shift={(#3\mytextsize,-#4\mytextsize)}]pic cs:end#5);
    \end{tikzpicture}}

\usepackage{hyperref} 
\hypersetup{colorlinks=true, citecolor=blue, urlcolor=blue, linkcolor=blue}
\usepackage{cleveref}
  	\crefname{figure}{Figure}{Figures}
  	\crefname{table}{Table}{Tables}
  	\crefname{equation}{Eq.}{Eqs.}
  	\crefname{section}{Section}{Sections}
  	\crefname{subsection}{Section}{Sections}
  	\crefname{subsubsection}{Section}{Sections}
  	\crefname{algorithm}{Algorithm}{Algorithms}

\usepackage[section]{minted}
\usepackage{newunicodechar}
\newunicodechar{δ}{\ensuremath{\delta}} 

\newunicodechar{μ}{\ensuremath{\mu}}
\newunicodechar{σ}{\ensuremath{\sigma}}
\newunicodechar{ν}{\ensuremath{\nu}}
\newunicodechar{λ}{\ensuremath{\lambda}}

\newcommand{\code}[1]{\texttt{#1}}
\newcommand{\bigO}[1]{\ensuremath{\mathcal{O}({#1})}}

\usepackage{todonotes}

\definecolor{hokiemaroon}{RGB}{134,31,65}
\definecolor{hokieorange}{RGB}{229,117,31}
\definecolor{hokiestone}{RGB}{117,120,123}

\newcommand{\atensor}[2]{#1\vphantom{|}\indices*{#2}}

\begin{document}


\title{\code{SeQuant} Framework for Symbolic and Numerical Tensor Algebra. I. Core Capabilities.} 

\author{Bimal Gaudel}
\affiliation{Department of Chemistry, Virginia Tech, Blacksburg, VA, USA}

\author{Robert G. Adam}
\affiliation{Institute for Theoretical Chemistry, University of Stuttgart, Germany}

\author{Ajay Melekamburath}

\author{Conner Masteran}

\author{Nakul Teke}

\author{Azam Besharatnik}
\affiliation{Department of Chemistry, Virginia Tech, Blacksburg, VA, USA}

\author{Andreas K\"ohn}
\affiliation{Institute for Theoretical Chemistry, University of Stuttgart, Germany}

\author{Edward F. Valeev}
\email{efv@vt.edu}
\affiliation{Department of Chemistry, Virginia Tech, Blacksburg, VA, USA}


\date{\today}

\begin{abstract}
\code{SeQuant} is an open-source library for symbolic algebra of tensors over commutative (scalar) and non-commutative (operator) rings. The key innovation supporting most of its functionality is a graph-theoretic tensor network (TN) canonicalizer that can handle tensor networks with symmetries faster than their standard group-theoretic counterparts. The TN canonicalizer is used for routine simplification of conventional tensor expressions, for optimizing application of Wick's theorem  (used to canonicalize products of tensors over operator fields), and for manipulation of the intermediate representation leading to the numerical evaluation. Notable features of \code{SeQuant} include support for noncovariant tensor networks (featuring hyperedges in their graphical representation and often arising from tensor decompositions) and for tensors with modes that depend parametrically on indices of other tensor modes (such dependencies between degrees of freedom are naturally viewed as nesting of tensors, ``tensors of tensors'' arising in block-wise data compressions in data science and modern quantum simulation). \code{SeQuant} blurs the line between pure symbolic manipulation/code generation and numerical evaluation by including compiler-like components to optimize and directly interpret tensor expressions using external numerical tensor algebra frameworks. The \code{SeQuant} source code is available at \url{https://github.com/ValeevGroup/SeQuant}.
\end{abstract}

\maketitle 

\section{Introduction}\label{sec:intro}

\code{SeQuant} is a software component whose development was originally motivated by the need to symbolically derive and manipulate equations of approximate simulation methods for quantum many-body problems, such as many-body perturbation theories, coupled-cluster (CC) methods,\cite{VRG:shavitt:2009:} and propagator methods,\cite{VRG:dickhoff:2008:} among others. Although such theories were developed mainly manually, often using diagrammatic techniques,\cite{VRG:feynman:1949:PR,VRG:goldstone:1957:PRSMPES,VRG:hugenholtz:1957:P,VRG:harris:1992:} the implementation of many such methods is impractical without computer assistance. In the domain of chemical physics, and, more specifically, electronic structure,
tools like \code{SeQuant} have been used since the 1970s\cite{VRG:paldus:1973:CPC} to realize methods that would not be practical otherwise, such as spin-adapted\cite{VRG:janssen:1991:TCA,VRG:herrmann:2022:JCP}
and high-order\cite{VRG:kallay:2001:JCP,VRG:hirata:2003:JPCA,VRG:baumgartner:2005:PI,VRG:engels-putzka:2011:JCP} single-reference coupled-cluster methods, explicitly correlated (F12) coupled-cluster,\cite{VRG:shiozaki:2008:PCCP,VRG:shiozaki:2008:JCP,VRG:kohn:2008:JCP,VRG:shiozaki:2009:JCP}
internally-contracted multireference coupled-cluster\cite{VRG:neuscamman:2009:JCP,VRG:hanauer:2011:JCP,VRG:datta:2011:JCP,VRG:black:2023:JCP},
and analytic gradients of internally-contracted multireference perturbation theory,\cite{VRG:macleod:2015:JCP}
among others.\cite{VRG:lyakh:2005:JCP,VRG:lesiuk:2020:JCTC,VRG:masteran:2023:JCP,VRG:teke:2024:JPCA}
Besides the development of new methods automated tools can also be used to discover and correct errors in the manual implementations of complex methods.\cite{VRG:yanai:2017:MP,VRG:masteran:2023:JCP}

Although the best design practices for tools like \code{SeQuant} were established by Janssen and Schaefer more than 35 years ago\cite{VRG:janssen:1991:TCA}
, there have been many ongoing efforts motivated by new use cases \cite{VRG:nooijen:2001:JMS,VRG:kallay:2001:JCP,VRG:hirata:2003:JPCA,VRG:bochevarov:2004:JCP,VRG:piecuch:2006:IJQC,VRG:shiozaki:2008:JCP,VRG:kohn:2008:JCP,VRG:kohn:2009:JCP,zitko2011,saitow2013,VRG:saitow:2015:JCTC,krupicka2017,VRG:zhao:2018:,VRG:arthuis:2018:CPC,VRG:lesiuk:2020:JCTC,VRG:rubin:2021:MP,VRG:evangelista:2022:JCP,VRG:quintero-monsebaiz:2023:AA,VRG:lechner:2024:PCCP,VRG:liebenthal:2025:JPCA,VRG:brandejs:2025:JCTC,VRG:lexander:2025:}. Is there a need for yet another such tool? The long-term answer is no. What is actually needed is a reusable\cite{VRG:lehtola:2023:JCP} community framework that provides the robust symbolic manipulation capabilities of these tools. Or perhaps these goals can be achieved by domain-specific extensions of the existing community tool \code{SymPy}\cite{VRG:meurer:2017:PCS} or by adopting an existing field-theoretic CAS (such as \code{Cadabra},\cite{VRG:peeters:2018:J} which also based on \code{SymPy}). Our goal for \code{SeQuant} is to seed such a community framework, or at least to illustrate the best practices and test-drive the necessary algorithms.
This manuscript is the first in a series of reports on the most recent (second) version of \code{SeQuant},
developed since 2018 to replace the general (but slow) \code{Mathematica}-based version 1 of \code{SeQuant}.
Its design objectives include:
\begin{itemize}
\item {\em new semantic features}, most importantly
\begin{itemize}
\item the nested index dependencies that appear in many-body methods based on pair-natural orbitals (PNOs) such as the DLPNO coupled-cluster \cite{VRG:pinski:2015:JCP,VRG:riplinger:2016:JCP}
\item support for {\em noncovariant} products of tensors that involve hyperedges in their graphical representation due to summation over three or more instances of a given index, Hadamard-product-like indices, and other more general products; such tensor network structures appear in DLPNO methods as well as in methods that involve high-order global tensor factorizations, such as canonical polyadic, pseudospectral, tensor hypercontraction, etc.\cite{VRG:hitchcock:1927:JMP,VRG:carroll:1970:P,VRG:harshman:1970:WPP,VRG:beylkin:2002:PNAS,VRG:pierce:2023:JCTC,VRG:friesner:1985:CPL,VRG:hohenstein:2012:JCP,VRG:pierce:2021:JCTC}
\item support for user-defined index space vocabularies that involve arbitrarily complicated index space nestings, multiple types of basis representations for each space (e.g., canonical vs localized), which are needed for example to develop local explicitly correlated (F12) methods\cite{VRG:tew:2011:JCP,VRG:pavosevic:2016:JCP,VRG:kumar:2020:JCP,VRG:ma:2017:JCTC} that can feature 3 or more atomic orbital bases and dozens of orbital spaces.\cite{VRG:masteran:2025:JCTC} 
\end{itemize}
\item {\em close-to-optimal efficiency} to allow symbolic manipulation during a compilation stage (e.g., during code generation) as well as during computational tasks on particular molecules ({\em online} execution) to permit problem-specific optimization,
\item support for direct {\em interpretation} of the symbolic expressions to avoid the need for code generation, and
\item {\em improved reusability} as a component in a native code as well as a backend for interpreted languages.
\end{itemize}

The development of \code{SeQuant} involved substantial algorithmic development. Notably, the support nested index dependencies and noncovariant products of tensors to the best of our knowledge is not found anywhere else (uses of symbolic automation in the context of PNO-based perturbation theory exist\cite{VRG:saitow:2020:JCP} but appear limited in scope and few details are available). Thus, the goal of this first manuscript is to primarily describe the new algorithmic innovations of \code{SeQuant} found in its most recent version, 2.2.0\cite{valeev_2026_18689273}.
Namely, to enable symbolic manipulation of these extended algebraic tensor structures, we introduced a method for symmetry-preserving encoding of tensor networks as a colored graph; the established techniques for colored graph automorphism group computation and canonicalization then lead to an efficient method for computing automorphisms and canonical forms of such networks.
By supporting complex index dependencies, noncovariant networks, and general symmetries, our approach goes beyond other graph-theoretic approaches used in symbolic tensor algebra\cite{VRG:obeid:2001:,VRG:bolotin:2013:,VRG:li:2017:P2AISSAC,VRG:peeters:2018:J,VRG:kryukov:2019:JPCS} and, as we demonstrate, the approach beats the performance of traditional group-theoretic canonicalizers for tensor network(TN) symmetries.
In \code{SeQuant} the graph-theoretic TN canonicalizer supports not only the basic tensor expression simplification, but also most other aspects of symbolic tensor algebra, such as efficient canonicalization of products of quantum fields (Wick's theorem), optimization and runtime interpretation of tensor expressions, etc. Thus, the focus of this paper is primarily on the core symbolic capabilities of \code{SeQuant}, including the TN canonicalizer and its uses. We also describe how \code{SeQuant} blends symbolic manipulations with compiler-like symbolic optimization and lowering to executable form, culminating in code generation or numerical interpretation of tensor expressions at runtime using an external  numerical tensor backend. Subsequent papers will demonstrate additional features of \code{SeQuant} designed for the needs of quantum many-body simulation in chemistry and materials and highlight the performance of numerical evaluation.

It is important to note that the algorithmic innovations at the heart of \code{SeQuant} are related to the developments of symbolic tensor algebra capabilities in other (classical and quantum) field-theoretic domains where the use of machine-assisted symbolic manipulation of mathematical expressions preceded the chemical physics domain by at least a decade.
Expressions involving tensors occurred as key targets for symbolic manipulation of classical\cite{VRG:fletcher:1967:AJ} and quantum\cite{VRG:wilcox:1961:,VRG:kaiser:1963:NP} field-theoretic simulations already in the 1960s, almost from the beginning of general-purpose symbolic computation itself\cite{10.1145/800005}.
Despite seemingly drastic differences between the classical and quantum field-theoretic contexts, tensor structure of expressions poses common challenges for symbolic manipulations in both domains.\cite{VRG:korolkova:2013:PCS}
Viewing tensors abstractly as symbolic objects with index slots, tensor expressions in physical simulation domains involve tensors with slot permutation symmetries (or obeying more complex multiterm identities) and joined into tensor networks by summations over common indices.
The key building blocks for symbolic computation on tensor expressions, such as finding canonical forms\cite{VRG:geddes:1992:} of tensors and tensor networks,
traditionally are not provided by general-purpose computer algebra systems (CAS) such as \code{Mathematica}, \code{Maple}, etc. Thus, most of symbolic computation in classical and quantum simulation contexts involves custom algorithms\cite{VRG:butler:1991:,VRG:portugal:1998:CPC,VRG:portugal:1999:JPMG,VRG:ilyin:2000:PCS,VRG:bolotin:2013:,VRG:niehoff:2018:CPC,VRG:price:2022:} to deal with tensor algebra, often embedded in purpose-built tools\cite{VRG:bauer:2002:JoSC,VRG:peeters:2007:CPC,VRG:bolotin:2013:,VRG:peeters:2018:J}. Since symbolic manipulation in both domains often involves traditional (``scalar'') algebra and calculus, and thus conventional CAS machinery for representing and manipulating mathematical expressions is needed, tensor algebra algorithms are often implemented as addons to generic CASs.\cite{VRG:ilyin:1996:CPC,VRG:bauer:2002:JoSC,VRG:manssur:2004:CPC,VRG:martin-garcia:2008:CPC,VRG:peeters:2007:CPC}
This is in contrast to the domain of quantum many-body simulation where full-featured CAS capabilities are typically not needed. Nevertheless, the need to deal with the tensor structure of the expressions is what is common to tools like tools like \code{SeQuant} and the aforementioned symbolic tools for field-theoretic domains.

\code{SeQuant} consists of the following {\em core} (domain-neutral) modules:
\begin{itemize}
\item \code{SQ/symb}: domain-neutral symbolic transformation layer, comprising two subcomponents:
\begin{itemize}
 \item \code{SQ/expr}: abstract algebraic expression layer
 \item \code{SQ/tensor}:  tensor algebra, including
  \begin{itemize}
  \item fast canonicalizer for general tensor networks, and
  \item fast Wick's theorem engine for canonicalizing tensor networks of tensor operators on Fock space;
  \end{itemize}
\end{itemize}
\item \code{SQ/eval}: numerical evaluation of \code{SeQuant} expressions;
\end{itemize}
The core modules, collectively referred to as \code{SQ/core}, are discussed in more detail in \cref{sec:symb,sec:eval}, respectively. \Cref{sec:summary} contains a brief summary.

\section{\code{SQ/symb}: symbolic tensor algebra}\label{sec:symb}

\code{SQ/symb} includes basic capabilities for representing and manipulating mathematical expressions involving scalars and tensors. \code{SQ/symb} is not a full-featured system for symbolic math; it only supports a tiny subset of symbolic math that is of use for tensor manipulation (e.g., \code{SQ/symb} knows nothing about solving algebraic equations, special functions, or calculus). There are many full-featured {\em Computer Algebra Systems} (CAS) that can be used for symbolic math; while traditionally available in the form of monolithic packages (\code{Mathematica}, \code{Maple}, \code{Sage}) recently full-featured CAS libraries have also appeared, most notably \code{SymPy}. Although the original \code{SeQuant} was implemented in \code{Mathematica} to avoid reinventing the wheel, it was clear that a much more performant symbolic core was needed for the new version of \code{SeQuant}. We briefly considered \code{symengine}, a C++ CAS library designed to serve as a possible performant backend for \code{SymPy}, but for a number of reasons we decided to develop a custom symbolic engine for \code{SeQuant}. In the following, we describe its essential features; for more details, the reader is referred to the \code{SeQuant} source code available at \url{https://github.com/ValeevGroup/SeQuant}.

\subsection{\code{SQ/expr}: Expressions}\label{sec:expr}
The foundation of \code{SQ/symb}, just like any CAS, is the abstract expression layer.
An {\em expression} (\code{Expr}) either has no subexpressions (i.e., it is an {\em atom}) or is a sequence of subexpressions.
Since it is traditional to represent expressions as trees, we speak of expressions having no {\em children} (an atom) or one or more children (a sequence).

The recursive nature of expressions is reflected in the recursion-oriented API of \code{Expr}. Specifically, every \code{Expr} is a C++ {\em range} (i.e., a sequence) of references to \code{ExprPtr}s; atoms are empty ranges. As C++ ranges \code{Expr} can be used with a variety of generic algorithms provided by the C++ standard library and by nonstandard range libraries (\code{range-v3}, \code{Boost.Range}). The ability to apply generic range algorithms keeps the list of specialized expression algorithms \code{SeQuant} to a minimum. Unfortunately, existing range libraries typically do not provide algorithms for recursive ranges with arbitrary recursion depth. Therefore, \code{SeQuant} fills this gap by providing \code{visit}, which recursively iterates over the ranges and applies a given callable to each encountered range (or optionally to atoms only). Visitation serves as the key building block for many other algorithms.

The expression layer core is sufficiently generic to support user-defined expression types and their manipulation, i.e., users can define their own specializations of \code{Expr} and the core algorithms like visitation will work on them.
This allows us to implement {\em domain-specific} languages (DSLs). As will be shown in the second paper, the atom type \code{mbpt::Operator} will be introduced to represent abstract Fock-space linear operators ($\hat{A}$, $\hat{B}$, ...) for a more efficient composition and simplification of their algebra, without referring to their internal many-body structure.

To support {\em algebraic} expressions \code{SeQuant} provides the following core expression atom types:
\begin{itemize}
\item \code{Constant}: a {\em scalar} value, such as $1, 2+ 3\,\text{i}, 2/3 + 4/5\, \text{i}$; \code{Constant}s are internally represented by complex numbers over arbitrary precision rationals.
\item \code{Variable}: a {\em symbol} representing a scalar ($\omega$, $\hbar$)
\item \code{Tensor}: a tensor of scalars (see \cref{sec:tensor:scalars-and-operators}).
\item \code{NormalOperator}: a tensor of normal-ordered operators (see \cref{sec:tensor:scalars-and-operators}).
\end{itemize}
\code{Tensor} and \code{NormalOperator} are specializations of the class  \code{AbstractTensor} that represents an abstract indexed symbolic tensor (see \cref{sec:tensor:abstract_tensor}).

Atoms can appear in the following composite algebraic expressions:
\begin{itemize}
\item \code{Sum}: associative sum of one or more expressions
\item \code{Product}: an associative product of one or more expressions, optionally multiplied by a scalar prefactor; multiplication is not necessarily commutative.
\end{itemize}

\begin{figure}[ht!]
    \centering
    \includegraphics[width=\linewidth]{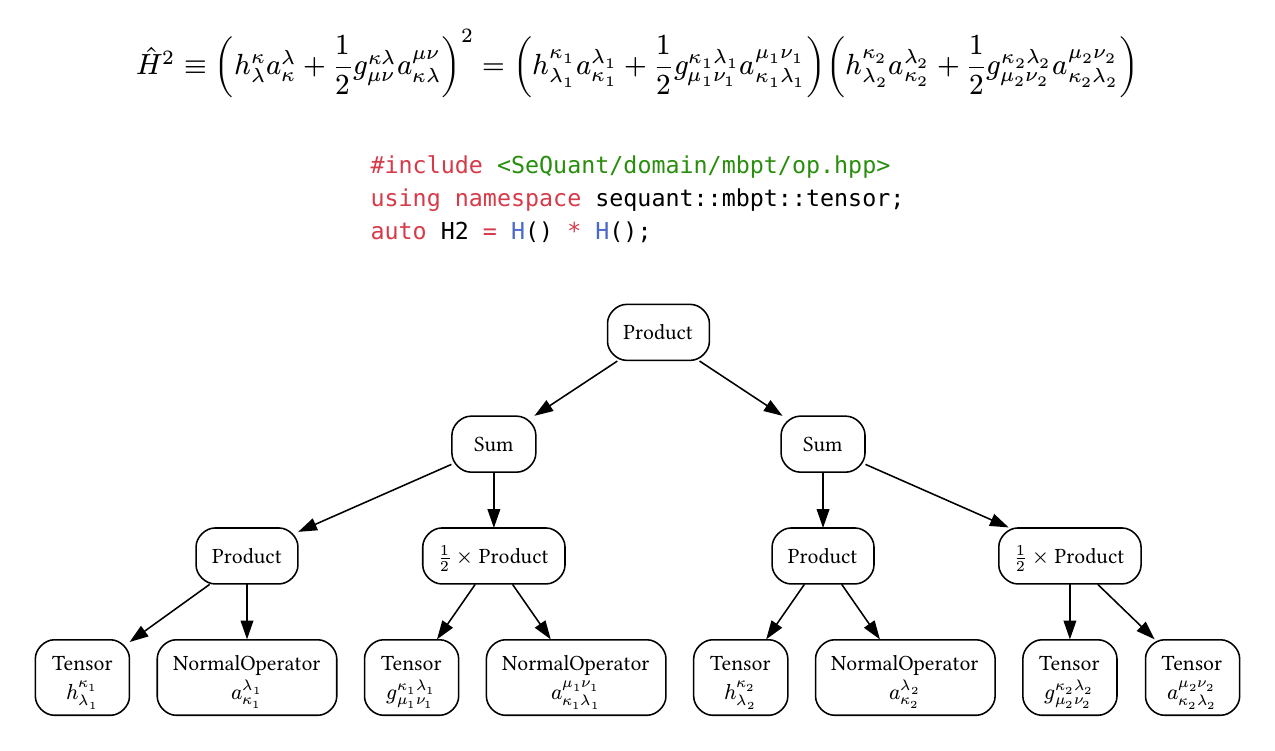}
    \caption{Programmatic construction and tree representation of a representative \code{SeQuant} expression.}
    \label{fig:expr0}
\end{figure}

\Cref{fig:expr0} illustrates how these expression types are used to represent the square of a Fock-space 2-body Hamiltonian. Here is a brief walkthrough of this example.
\begin{itemize}
\item Just like in the formal equations in this paper, Einstein's summation convention is used for \code{SeQuant} expressions involving \code{AbstractTensor} objects, i.e., summation is implied for all indices that appear on two different tensors within a \code{Product} of two or more objects that contain \code{AbstractTensor} objects.
\item The square of the Hamiltonian is a \code{Product} of two of its instances. Due to the implicit summation, in each instance of the Hamiltonian operator $\hat{H}$ a fresh set of dummy indices must be used to avoid ambiguities. Hence abstract operators, such as $\hat{H}$, are represented by {\em callables} (C++ functions, lambdas), calling which (\code{H()}) produces their concrete instances. Each instance of the Hamiltonian is represented as a \code{Sum} of two \code{Product}s, each containing the 1- and 2-body contributions, respectively.
\end{itemize}

In addition to the core abstract expression operations (such as visitation), \code{SeQuant} provides a minimal set of algorithms for rewriting algebraic expressions:
\begin{itemize}
\item \code{expand}: expands products of sums;
\item \code{canonicalize}: puts expressions into their {\em canonical} form, such as by sorting the summands in \code{Sum} and factors in \code{Product} (modulo commutativity);
\item \code{rapid\_simplify}: for performing trivial simplifications, such as eliminating trivial operations (multiplication by 1, addition of 0, etc.); many such simplifications are already performed by the constructors of \code{Sum} and \code{Product}.
\end{itemize}
These operations can be combined nontrivially, among themselves, and with C++ range algorithms.

For convenience \code{SeQuant} provides a top-level \code{simplify} function to produce a ``simple'' expression by a particular combination of such transformations. But since there is no single definition of {\em simple} the user may have to do sequences of manipulations to put the expression into the desired form.

\Cref{listing:visit} illustrates how visitation can be used to traverse and rewrite expressions, and then combined with algebraic simplification (for brevity code listings omit \code{\#include} macros and explicit namespaces; the reader is referred to the \code{SeQuant} unit tests for the full-fledged form of the code listings contained in this manuscript).
The body of the visitor in \cref{listing:visit} follows the typical pattern-matching structure: if subexpression matches pattern 1 do action 1, else try pattern 2, etc. Since the visitor receives an abstract subexpression whose type is only known at runtime, pattern matching starts by querying the type of the subexpression.

\begin{widetext}
\begin{center}
\begin{listing}[H]
\begin{minted}{c++}
// x = (1 + 2) * (3 + 4)
auto x = (ex<Constant>(1) + ex<Constant>(2)) *
         (ex<Constant>(3) + ex<Constant>(4));

// scales every Constant in this expression by 2, i.e.
// converts x to (2 + 4) * (6 + 8)
x->visit([](ExprPtr& subexpr) {
  // every Constant subexpression ...
  if (subexpr->is<Constant>()) {
    // ... is replaced by its doubled counterpart
    subexpr = ex<Constant>(2 * subexpr->as<Constant>().value());
  }
});

// simplification attempts a variety of expression rewrites,
// including carrying out trivial arithmetic which makes x = 84
simplify(x);
assert(x->is<Constant>());
assert(x->as<Constant>().value() == 84);
\end{minted}
\caption{Example of rewriting a \code{SeQuant} expression using a visitor.}
\label{listing:visit}
\end{listing}
\end{center}
\end{widetext}

All expression objects are dynamically allocated and accessed by managed pointers (\code{ExprPtr}). This makes many types of expression rewriting, such as algebraic simplification, much faster by removing the need to rebuild the entire expression tree to make a localized change.

\subsection{\code{SQ/tensor}: Tensors}\label{sec:tensor}

\subsubsection{\code{SQ/tensor}: \code{AbstractTensor}}\label{sec:tensor:abstract_tensor}

\code{SeQuant} was designed to manipulate specific types of tensor algebras, involving conventional tensors of scalars, tensors of operators (ubiquitous in the quantum field-theoretic context), and their further generalizations.
Due to the sheer variety of types of tensors that \code{SeQuant} can encounter, let us start with a brief glossary. The notion of a tensor appears in a wide variety of contexts as a representation of a multivariate entity.
In some contexts, such as data science, tensors are just multidimensional arrays.
In physical simulation contexts, properties of tensors under a variety of symmetry transformations (coordinate rotations, particle interchanges, time reversal, etc.) 
must be reflected in their structure, thereby leading to more complex definitions.
Tensor of order-$(p,q)$ (with $p$ contravariant and $q$ covariant {\em modes}) on $p+q$ vector spaces $\{V_i: i=1\ldots p+q\}$ is a multilinear map from a product of $p$ dual vector spaces $\{V^*_i: i=1\ldots q\}$
and $q$ primal vector spaces $\{V_i: i=p+1\ldots p+q\}$ to a {\em ring} $R$:\footnote{Tensors are commonly defined over fields. To support manipulation of tensors of operators \code{SeQuant} restricts dealing with tensors over (commuting or non-commuting) rings only (see \cref{sec:tensor:scalars-and-operators}).}
\begin{align}
\label{eq:T-pq}
T: V_1^* \times \ldots \times V_p^* \times V_{p+1} \times \ldots \times V_{p+q} \rightarrow R
\end{align}
The products of vector spaces can have additional structure; for example, the use of Grassman's wedge product results in tensors totally antisymmetric with respect to permutation within covariant and within contravariant components, which is needed for simulation of collections of identical fermions. In such a case, all contravariant and covariant modes must refer to the same vector spaces (e.g., $V_1$ and $V_2$):
\begin{align}
\bar{T}: V_1^* \wedge \ldots \wedge V_1^* \times V_{2} \wedge \ldots \wedge V_{2} \rightarrow R
\end{align}
More general symmetries are needed in practice (see below).

The choice of a specific basis $\{v^{(i)}_k\}$ to represent each vector space $V_i$ leads to a specific {\em representation} of tensor $T$ as an array of values:
\begin{align}
    T^{k_1 \ldots k_p}_{k_{p+1} \ldots k_{p+q}} \equiv T[v^{(1)}_{k_1}, \ldots v^{(p+q)}_{k_{p+q}}],
\end{align}
where the {\em super}script and {\em sub}script indices index the basis sets of the {\em contra}variant and {\em co}variant modes, respectively. Although this {\em cosub} convention is nearly universal, in some contexts the opposite, {\em contrasub}, convention is used.\cite{VRG:harris:1981:PRA,VRG:monkhorst:1981:PRA,VRG:jeziorski:1981:PRA,VRG:kutzelnigg:1982:JCP}
In this work all equations use the latter convention. 
In most contexts we will avoid terms contravariant and covariant altogether in lieu of their shorter {\em bra} and {\em ket} counterparts borrowed from the Dirac bra-ket notation\cite{VRG:dirac:1939:MPCPS}.

Numerical computations involving tensors always involve a specific choice of basis; hence, numerical tensor libraries represent tensors as multidimensional arrays of values that correspond to a specific choice of basis for each tensor mode.
Thus, using index notation to refer to numerical elements of tensor representations is natural. However, even in purely symbolic manipulation of tensor algebras, explicit use of indices is convenient. {\em Abstract}
index notation\cite{VRG:penrose:1971:Andt} in which a symbolic index is bound to each mode is a powerful way to encode the tensor expressions without the loss of generality due to the choice of a specific basis.
It is natural as a way to compose and represent tensor networks in software.\cite{VRG:fishman:2022:SPC}
In \code{SeQuant} class \code{AbstractTensor} is a model of such an abstract view.
The \code{AbstractTensor} representation of the order-$(p,q)$ tensor (\cref{eq:T-pq}) assigns unique \code{Index} objects to each of the $p$ bra and $q$ ket modes. The indices are then used to compose operations such as products of tensors, as discussed below. \code{AbstractTensor}'s \code{bra} and \code{ket} methods provide access to the sequences of bra and ket indices. The general model of a tensor requires distinguishing the notion of tensor indices, which are always associated with specific modes, from the notion of {\em slots}, which may be empty (see below). The empty slots are denoted by null \code{Index} objects in the bra/ket index sequences.

\code{AbstractTensor} also encodes various tensor symmetries:
\begin{itemize}
\item \code{Symmetry} with respect to the permutation of indices within bra or within ket,
\item \code{BraKetSymmetry} with respect to the permutation of bra and ket (this only makes sense if $p=q$ and the bra and ket vector spaces are duals of each other), and
\item \code{ColumnSymmetry} with respect to the permutation of matching pairs of bra/ket slots.
\end{itemize}

However, even in the context of physical simulation, it is useful to introduce the notion of tensor modes that do not refer to vector spaces. For example, when computing the response of a system with respect to multiple distinct perturbations at a time, it is useful to add a mode to the perturbation operator that refers to the perturbation index. Another example is provided by various high-order tensor decompositions, such as the canonical polyadic decomposition\cite{VRG:hitchcock:1927:JMP,VRG:carroll:1970:P,VRG:harshman:1970:WPP}, in which tensor of order $n$ is decomposed into $n$ (generally, non-unitary) matrices analogously to singular value decomposition:
\begin{widetext}
\begin{align}
T^{k_1 \ldots k_p}_{k_{p+1} \ldots k_{p+q}} \overset{\text{CP}}{=} \sum_r \lambda_r (M^{(1)})^{k_1}_r \dots (M^{(p)})^{k_p}_r (M^{(p+1)})_{k_{p+1}}^r \dots (M^{(p+q)})_{k_{p+q}}^r,
\end{align}
\end{widetext}
where $r$ indexes the ``rank'' mode of CP singular values $\lambda$ and factors $U^{(i)}$.
Such a decomposition (exact or approximate) is not invariant with respect to unitary transformation of the ``rank'' mode; hence it is not necessary to treat the rank mode as referring to a vector space.
To support non-vector-space modes \code{AbstractTensor} introduced another type of tensor mode: an array-like, or {\em aux}, mode. Thus, \code{AbstractTensor} can represent tensors, arrays, and their hybrids.
The standard array notation is used for the aux indices. For example, the element of a tensor with 1 bra, 1 ket, and 1 aux index is denoted $B^p_q[r]$

\subsubsection{\code{SQ/tensor}: Tensors of Scalars and Operators}\label{sec:tensor:scalars-and-operators}

Class \code{Tensor} specializes \code{AbstractTensor} to represent conventional tensors whose elements are scalars. Since a product of scalars is commutative, tensor elements can be reordered within products of tensors. For example, the product of matrices ${\bf A}$ and ${\bf B}$ (i.e., order-$(1,1)$ tensors) is not generally commutative:
\begin{align}
  A_p^r B_r^q \equiv ({\bf AB})_p^q \neq ({\bf BA})_p^q \equiv B_p^r A_r^q.
\end{align}
However, it is possible to reorder the matrix elements in the symbolic specification of product ${\bf AB}$:
\begin{align}
\label{eq:Apr-Brq}
  A_p^r B_r^q = B_r^q A_p^r.
\end{align}
Note that both sides of \cref{eq:Apr-Brq} denote the same matrix element $({\bf AB})_p^q$. The freedom to reorder the elements of the scalar tensor in the specification of the tensor network is due to the commutativity of the product of the scalars (in this case, $A_p^r$ and $B_r^q$).
Due to this freedom, tensor networks containing tensors of scalars (or any commutative rings) lack a unique representation; how to rewrite them in a unique canonical form will be discussed in \cref{sec:canon}.

In the context of quantum field theories we encounter tensors over noncommutative rings, such as tensors of {\em operators} built from creation and annihilation operators. For example, x order-$(1,1)$ tensors composed of {\em fermionic} annihilation $a_p$ and creation $a^p \equiv \left(a_p\right)^\dagger = a^\dagger_p$ operators that satisfy canonical anticommutation relations (CARs):
\begin{align}
\label{eq:CAR-aa}
[a_p, a_q]_+ = & 0 \\
\label{eq:CAR-cc}
[a^p, a^q]_+ = & 0 \\
\label{eq:CAR-ca}
[a^p, a_q]_+ = & \delta^p_q
\end{align}

where $\delta^p_q$ is the Kronecker delta. Note that for the case of nonorthogonal bra (ket) sp states, it may be necessary to distinguish the notion of covariance/contravariance from bra/ket\cite{VRG:burton:2021:JCP} but here we assume mutually orthonormal bra (ket) sp states that may or may not be biorthogonal. For non-biorthogonal sp bra/ket states, the Kronecker delta is replaced by the metric tensor $s^p_q \equiv \braket{q}{p}$.
Consider a tensor operator of order $(1,1)$:
\begin{align}
    C_p^q \equiv a^p a_q.
\end{align}
\cref{eq:Apr-Brq} no longer holds with $C$ replacing $A$ and $B$, as can be easily shown by the application of CARs:
\begin{align}
\label{eq:Cqr-rp}
 C_r^q C_p^r \equiv & a^q a_r a^r a_p = - a^q a^r a_r a_p \\
 C_p^r C_r^q \equiv & a^r a_p a^q a_r = a^r (\delta^q_p - a^q a_p) a_r = \delta^q_p a^r a_r - a^q a^r a_r a_p \nonumber \\
 \overset{\rm \cref{eq:Cqr-rp}}{=} &  \delta^q_p C^r_r + C_r^q C_p^r \neq C_r^q C_p^r.
\end{align}
The loss of freedom to reorder the tensor elements is due to the noncommutative multiplication of elements of $C$.
Canonicalization of tensor networks containing noncommutative tensor (e.g., operator) rings will have to account for the lack of reordering freedom.

Class \code{NormalOperator} specializes \code{AbstractTensor} to represent tensors whose elements are products of fermionic/bosonic creation/annihilation operators that are {\em normal-ordered} with respect to a vacuum. For example, a normal-ordered fermionic operator composed of $n_c$ creators and $n_a$ annihilators is denoted by\cite{VRG:kutzelnigg:1982:JCP}
\begin{align}
  a^{p_1 \dots p_c}_{q_1 \dots q_a} \equiv \{ a^{p_1} \dots a^{p_c} a_{q_a} \dots a_{q_1} \},
\end{align}
where $\{ \ldots \}$ symbolizes normal ordering;
\code{SeQuant} supports normal ordering with respect to genuine vacuum (for fermions and bosons) and single-configuration (Fermi) vacuum (for fermions only, denoted by $\tilde{a}$ to distinguish from genuine normal-ordered operators denoted by $a$). 
The annihilation/creation indices of \code{NormalOperator} correspond to the bra/ket indices of \code{AbstractTensor}, respectively. Due to the creators and annihilators of fermions/bosons anticommuting/commuting, the bras and kets are antisymmetric/symmetric, respectively.

\subsubsection{\code{SQ/tensor}: Index Dependencies}\label{sec:tensor:index_dependencies}

In some reduced-complexity quantum simulation methods, tensors are compressed in a slicewise fashion.\cite{VRG:neese:2009:JCPa,VRG:yang:2011:JCP,VRG:pinski:2015:JCP} That is, tensor slices corresponding to fixed values of indices of a subset of modes are compressed in some fashion. For example, consider the tensor of order (2,2), $t^{i_1 i_2}_{a_1 a_2}$. Slice $i_1 i_2$ is the set of all elements $t^{i_1 i_2}_{a_1 a_2}$ with fixed $i_1 i_2$; it can be viewed as a matrix with elements indexed by $a_1$ and $a_2$:
\begin{align}
({\bf t}^{i_1 i_2})_{a_1 a_2} \equiv t^{i_1 i_2}_{a_1 a_2}.
\end{align}
Approximate singular value decomposition (SVD) of matrix ${\bf t}^{i_1 i_2}$,
\begin{align}
\label{eq:t-i1i2-svd}
    {\bf t}^{i_1 i_2} \approx {\bf U}^{i_1 i_2} {\bf \Sigma}^{i_1 i_2} ({\bf V}^{i_1 i_2})^\dagger
\end{align}
truncated to keep singular values not less than some threshold often reveals a low-rank structure that is not easily accessible to global decompositions.
Since the SVD rank varies with $i_1 i_2$, it is not possible to represent the aggregates of matrices ${\bf U}^{i_1 i_2}$ (or ${\bf V}^{i_1 i_2}$) as tensors of scalars. Probably the most natural representation of such an aggregate is an order-2 tensor (indexed by $i_1 i_2$) of matrices ${\bf U}^{i_1 i_2}$, i.e., a tensor of tensors. However, other representations are possible (e.g., as a sparse tensor of scalars). However, the details of numerical representation are not important for symbolic manipulations. What is essential is to express the idea that the column (singular) mode of ${\bf U}^{i_1 i_2}$ represents a vector space specific to each $i_1 i_2$; we say that $i_1 i_2$ is a {\em protoindex bundle} for such vector space.

The indices that refer to a vector space equipped with protoindices must also be annotated by the protoindex bundle. To support such index relationships each \code{Index} object in \code{SeQuant} can have zero or more protoindices. For example, an index referring to the column mode of ${\bf U}^{i_1 i_2}$ (\cref{eq:t-i1i2-svd})
will have 2 protoindices, $i_1$ and $i_2$; in mathematical expressions protoindices will be typeset in superscript, e.g., as $a_1^{i_1 i_2}$. Rewriting \cref{eq:t-i1i2-svd} in terms of such indices produces
\begin{align}
\label{eq:t-i1i2-a1a2-svd}
t^{i_1 i_2}_{a_1 a_2} \approx \atensor{U}{*^{a_3^{i_1 i_2}}_{}*^{i_1}_{a_1}*^{i_2}_{}} ~ \Sigma_{a_3^{i_1 i_2} a_4^{i_1 i_2}} ~ \left( \atensor{V}{*^{a_4^{i_1 i_2}}_{}*^{i_1}_{}*^{i_2}_{a_2}}\right)^*.
\end{align}
Note the alignment of bra/ket index pairs $\{i_1,a_1\}$ and $\{i_2,a_2\}$ on both left- and right-hand sides of \cref{eq:t-i1i2-a1a2-svd} and the resulting gaps (empty slots) in the tensors. Preserving the slot {\em bundles} (see below) such as these pairs of matching bra/ket slots is important for some manipulations and for reflecting certain symmetries (e.g., in the electronic structure context of this example\cite{VRG:pinski:2015:JCP} the left- and right-hand sides are symmetric with respect to exchange of index pairs $\{i_1,a_1\}$ and $\{i_2,a_2\}$).

As already mentioned, tensors with such index dependencies lack a unique numerical representation. But they also lack unique symbolic representation!
For example, $\atensor{U}{*^{a_3^{i_1 i_2}}_{}*^{i_1}_{a_1}*^{i_2}_{}}$ can be represented as $\atensor{U}{*^{a_3^{i_1 i_2}}_{a_1}}$ without loss of generality. In fact, such {\em compact} representation is more natural; whereas \cref{eq:t-i1i2-a1a2-svd} hardly resembles the standard SVD, 
\begin{align}
  \label{eq:t-i1i2-a1a2-svd-compact}
  t^{i_1 i_2}_{a_1 a_2} \approx \atensor{U}{*^{a_3^{i_1 i_2}}_{a_1}} \Sigma_{a_3^{i_1 i_2} a_4^{i_1 i_2}} \left( \atensor{V}{*^{a_4^{i_1 i_2}}_{a_2}}\right)^*
\end{align}
looks very much like matrix SVD. Whereas the {\em verbose} form, $\atensor{U}{*^{a_3^{i_1 i_2}}_{}*^{i_1}_{a_1}*^{i_2}_{}}$, makes it explicit that the tensor $U$ has 4 modes, 3 independent (indexed by $i_1$, $i_2$, and $a_1$) and 1 dependent (indexed by $a_3^{i_1 i_2}$), more importantly, it allows to specify symmetries more precisely. Namely, if $t^{i_1 i_2}_{a_1 a_2}$ is symmetric with respect to swapping of ``columns'' $\{ i_1, a_1\}$ and $\{ i_2, a_2\}$, the tensor $U$ must be related to the tensor $V$ as
\begin{align}
\atensor{U}{*^{a_3^{i_1 i_2}}_{}*^{i_1}_{a_1}*^{i_2}_{}} = \left(\atensor{V}{*^{a_3^{i_2 i_1}}_{}*^{i_1}_{} *^{i_2}_{a_1}} \right)^*.
\end{align}
However, the compact form is often more natural, as shown for the SVD example, and arises naturally. For example, the analog of CAR for fermionic creation/annihilation operators (\cref{eq:CAR-ca})
expressed in the pair-specific basis defined by the tensors $U$/$V$ in \cref{eq:t-i1i2-a1a2-svd} reads:
\begin{align}
\label{eq:CAR-pno}
  [a^{a_1^{i_1 i_2}}, a_{a_2^{i_3 i_4}}]_+ = & s^{a_1^{i_1 i_2}}_{a_2^{i_3 i_4}},
\end{align}
where the right-hand side instead of the expected Kronecker delta contains the inner product of the basis states for different pairs (such states are not orthonormal unless $i_1i_2 = i_3i_4$). The metric tensor $s^{a_1^{i_1 i_2}}_{a_2^{i_3 i_4}}$ expressed in the verbose notation would be a tensor with 6 modes indexed by $i_1$, $i_2$, $i_3$, $i_4$, $a_1^{i_1 i_2}$, and $a_2^{i_3 i_4}$. The compact notation is the clear winner.

\subsection{\code{SQ/tensor}: Tensor Network Canonicalizer}\label{sec:canon}

\subsubsection{Problem Statement}
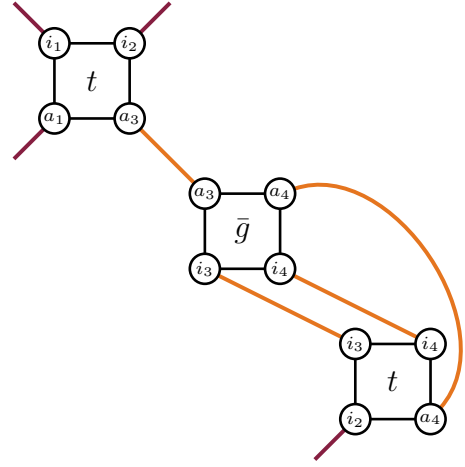
\begin{figure}
\begin{tikzpicture}
\SetVertexStyle[FillColor=white]
\Vertex[x=-2,y=2,label=$t$,style={draw,rectangle},size=1,fontsize=\large]{TL}
\Vertex[x=-2.5,y=2.5,size=.4,label=$i_1$]{TLnw}
\Vertex[x=-1.5,y=2.5,size=.4,label=$i_2$]{TLne}
\Vertex[x=-2.5,y=1.5,size=.4,label=$a_1$]{TLsw}
\Vertex[x=-1.5,y=1.5,size=.4,label=$a_3$]{TLse}
\Vertex[x=0,y=0,label=$\bar{g}$,style={draw,rectangle},size=1,fontsize=\large]{G}
\Vertex[x=-.5,y=.5,size=.4,label=$a_3$]{Gnw}
\Vertex[x=.5,y=.5,size=.4,label=$a_4$]{Gne}
\Vertex[x=-.5,y=-.5,size=.4,label=$i_3$]{Gsw}
\Vertex[x=.5,y=-.5,size=.4,label=$i_4$]{Gse}
\Vertex[x=2,y=-2,label=$t$,style={draw,rectangle},size=1,fontsize=\large]{TR}
\Vertex[x=1.5,y=-1.5,size=.4,label=$i_3$]{TRnw}
\Vertex[x=2.5,y=-1.5,size=.4,label=$i_4$]{TRne}
\Vertex[x=1.5,y=-2.5,size=.4,label=$a_2$]{TRsw}
\Vertex[x=2.5,y=-2.5,size=.4,label=$a_4$]{TRse}
\Edge[color=hokieorange](TLse)(Gnw)
\Edge[color=hokieorange](Gsw)(TRnw)
\Edge[color=hokieorange](Gse)(TRne)
\Edge[color=hokieorange,bend=75](Gne)(TRse)
\Vertex[x=-3.25,y=3.25,Pseudo]{TLnwp}
\Edge[color=hokiemaroon](TLnw)(TLnwp)
\Vertex[x=-0.75,y=3.25,Pseudo]{TLnep}
\Edge[color=hokiemaroon](TLne)(TLnep)
\Vertex[x=-3.25,y=0.75,Pseudo]{TLswp}
\Edge[color=hokiemaroon](TLsw)(TLswp)
\Vertex[x=0.75,y=-3.25,Pseudo]{TRswp}
\Edge[color=hokiemaroon](TRsw)(TRswp)
\end{tikzpicture}
\caption{Graphical representation of TN in \cref{eq:tn1}. External edges are shown in maroon, internal edges are shown in orange.}
      \label{fig:tn1}
\end{figure}

A key building block for many algorithms in symbolic tensor algebra involves rewriting tensor networks in their canonical forms by applying {\em symbolic} transformations that leave the tensor network invariant. Symbolic transformations exclude the more general mathematical transformations, such as gauge transformations; specialized canonicalization problems arise also in such contexts\cite{VRG:acuaviva:2023:2I6ASFCSF}. To understand what we mean by symbolic transformations, consider a concrete example from the coupled-cluster doubles (CCD)\cite{VRG:cizek:1966:JCP}
 equations:
\begin{align}
\label{eq:tn1}
r^{i_1 i_2}_{a_1 a_2} \equiv & \,
\bar{g}^{{a_3}{a_4}}_{{i_3}{i_4}} t^{{i_1}{i_2}}_{{a_1}{a_3}} t^{{i_3}{i_4}}_{{a_2}{a_4}},
\end{align}
where indices $\{i_k,a_k\}$ represent single-particle states \{occupied, unoccupied\} in the CC reference wave function, respectively.
The TN in \cref{eq:tn1} has 4 {\em external} indices, $i_1, i_2, a_1, a_2$, that are not summed over and thus appear on both sides of \cref{eq:tn1} and 4 {\em internal} indices, $i_3, i_4, a_3, a_4$, that are summed out and appear only on the right-hand side of \cref{eq:tn1}.\footnote{Not to be confused with other uses of ``external'' and ``internal'' indices in the electronic structure literature.\cite{VRG:schutz:2003:PCCP, VRG:janowski:2008:JCTC}} The graphical illustration of TN \cref{eq:tn1} (see \cref{fig:tn1}) makes the meaning of external/internal terms obvious: the external indices label edges that stick out of the tensor network, whereas the internal indices label edges confined within the tensor network.  The internal indices are also referred to as {\em dummy} indices since they can be renamed arbitrarily on the right hand side.\footnote{A renaming must preserve semantics to be valid, e.g., $i_3 \to i_9$ is valid, but $i_3 \to a_7$ is not.} The external indices in \cref{eq:tn1} are also arbitrary, but unlike the dummy indices cannot be renamed on the right-hand side only; sometimes we'll refer to such indices as {\em named} to indicate that they cannot be renamed in context of a particular operation.

Relabelings of dummy indices that do not introduce new indices to the tensor network can be viewed as index permutations; dummy index swaps $i_3 \leftrightarrow i_4$ and $a_3 \leftrightarrow a_4$ of \cref{eq:tn1} produce
\begin{align}
\label{eq:tn1-v2}
r^{i_1 i_2}_{a_1 a_2} = & \,
\bar{g}^{{a_4}{a_3}}_{{i_4}{i_3}} t^{{i_1}{i_2}}_{{a_1}{a_4}} t^{{i_4}{i_3}}_{{a_2}{a_3}},
\end{align}
which is a different {\em representation} of the same TN as \cref{eq:tn1}.
Such dummy index renamings are particular examples of TN {\em symmetries} that leave the TN {\em essentially} (modulo phase) unchanged. Clearly, index-based TN encoding is generally not unique.

The lack of uniqueness of index-based TN encoding is a problem that has had a prominent spotlight since the first attempts to use computerized tensor algebra.\cite{VRG:bogen:1977:LMP,VRG:hornfeldt:1979:SaAC}
It is reasonable to wonder if some alternative TN representations avoid this issue altogether.
For example, some tensor-centric domains, like tensor analysis and machine learning,\cite{VRG:kolda:2009:SR} traditionally prefer index-free notation that omits the appearance of bias to a particular basis set/numerical representation (clearly, a more abstract view of indexing as a composition/representation device avoids such bias also\cite{VRG:penrose:1971:Andt}) and raises the level of abstraction; for example, the (column-wise) Khatri-Rao product\cite{VRG:khatri:1968:SIJSS1} of matrices $\mathbf{A}$ and $\mathbf{B}$ is an order-3 tensor $\mathbf{A} \odot \mathbf{B}$  that can be represented compactly using index-based notation as follows:
\begin{align}
    \left(\mathbf{A} \odot \mathbf{B}\right)_{pqr} \equiv & A_{pr} B_{qr}.
\end{align}
Clearly, even for matrices the number and variety of products is already unwieldy\cite{VRG:liu:2008:IJISS} and such representation is not generally-applicable to tensors.
Another alternative is provided by diagrammatic representation of tensor networks, which is again dominant in physics\cite{VRG:feynman:1949:PR,VRG:penrose:1971:Andt} due to its amenability to manual (computer-free) manipulation. Computer representation of diagrammatically-encoded TNs will be similar/identical to that of index-encoded TNs. As will become clear from our approach to canonicalization of index-encoded tensor networks, graph-theoretic approaches effectively bridge/remove the gap between the diagrammatic and index TN encodings.

The nonuniqueness of the index-based encoding of TNs means that their symbolic manipulation requires converting TNs to their unique canonical representation. Finding a unique canonical TN for a given TN is known as the TN {\em canonicalization} problem; it is also known simply as the {\em tensor} canonicalization problem (since a TN can be viewed as a single tensor obtained as an outer product of the TN factors and partially or fully traced-out indices), or even as {\em index} canonicalization.\cite{VRG:martin-garcia:2008:CPC} The canonicalization problem is related to the {\em isomorphism-detection} or {\em matching} problem, i.e., deciding whether two TNs are equivalent, modulo ``reindexing''; the TN matching can be solved by canonicalization, albeit less efficiently in some cases. Since we have more general use cases than simple matching of equivalent TNs, isomorphism detection is not sufficient for our purposes. Hence, the canonicalization problem will be the only focus here.

To discuss the canonicalization problem further, we need to introduce the notion of the {\em automorphism} group of the tensor network.
The dummy index permutations discussed above form a subgroup of the full automorphism group. The automorphism group of TN in \cref{eq:tn1} involves all permutations of tensor {\em slots} that leave the tensor network invariant (modulo a trivial transformation, such as multiplication by -1 or complex conjugation). A slot is a placeholder for an index; e.g., tensor element $t^{{i_1}{i_2}}_{{a_1}{a_3}}$
has 4 slots, two ket (typed as superscript), occupied by $i_1$ and $i_2$, respectively, and two bra (typed as subscript), occupied by $a_1$ and $a_2$, respectively. The network in \cref{eq:tn1} has 12 slots (6 bra and 6 ket). Each internal index occupies two slots (one bra and one ket), whereas each external index occupies one.
Some slots may be empty; see for example the tensor $U$ in \cref{eq:t-i1i2-a1a2-svd}. Protoindex bundles can also be viewed as sequences of slots.

Slots form logical groups, or slot {\em bundles}. We already encountered 4 such bundles: sequences of bra, ket, and aux slots, as well as the protoindex bundles. Another important bundle is a pair of matching bra/ket slots placed in a column (e.g., slots occupied by $i_1$ and $a_1$ in $r^{i_1 i_2}_{a_1 a_2}$); we will refer to such bundles as {\em column} bundles. Column bundles can include empty slots, such as the first column bundle of tensor $U$ in \cref{eq:t-i1i2-a1a2-svd}. Bundles can be nested; e.g., for tensors with totally symmetric/antisymmetric bra and ket we will denote the pair of bra and ket bundles as {\em braket} bundle. For example, tensor $\atensor{r}{*^{i_1}_{a^{i_1 i_2}_1}*^{i_2}_{a_2^{i_1 i_2}}}$ has the following bundles:
\begin{itemize}
    \item 1 bra bundle: $a^{i_1 i_2}_1 a^{i_1 i_2}_2$,
    \item 1 ket bundle: $i_1 i_2$,
    \item 2 column bundles: $i_1 a^{i_1 i_2}_1$ and $i_2 a_2^{i_1 i_2}$,
    \item 1 braket bundle consisting of the bra and the ket bundles,
    \item 2 protoindex bundles: $i_1 i_2$ in $a^{i_1 i_2}_1$ and  $i_1 i_2$ in $a^{i_1 i_2}_2$.
\end{itemize}

In the physics/chemistry literature the notion of slots is often left implicit, even when discussing tensor-algebraic expressions with explicit indexing such as \cref{eq:tn1}. This makes discussion of symmetries of tensor networks (rather than individual tensors) inconvenient since an index can appear more than once in an expression, hence it can occupy multiple slots; for example, index $a_3$ in \cref{eq:tn1} occupies 2 slots. In covariant tensor notation, compact slot specification by superscript/subscript index is still possible (e.g., index $a_3$ in \cref{eq:tn1} occupies slots $^{a_3}$ and $_{a_3}$) but there are increasingly many uses of {\em noncovariant} tensor networks in which each index can appear more than twice, thereby making the explicit notion of slots important.

Most tensor-algebraic contexts in practice involve tensor networks that are {\em slot-symmetric}, i.e., they are invariant (up to a phase factor) with respect to one or more slot permutations.
The tensor network can remain invariant under permutations of slots for several reasons:
\begin{enumerate}
    \item {\bf Tensor Symmetries}. Invariance of the constituent tensors with respect to the permutations of its equivalent {\em modes} (if any). For example, tensors $\bar{g}$ and $t$ in \cref{eq:tn1} are antisymmetric with respect to permutations of their bra/ket indices, e.g.:
    \begin{align}
        t^{{i_1}{i_2}}_{{a_1}{a_3}} = -t^{{i_2} {i_1}}_{{a_1}{a_3}} = t^{{i_2} {i_1}}_{{a_3} {a_1}} = -t^{{i_1} {i_2}}_{{a_3}{a_1}}.
    \end{align}
    Hence, the following primitive permutations (transpositions) of slots,
\begin{align}
\hat{P}^{a_3,a_4} \equiv \, & ^{a_3} \leftrightarrow {}^{a_4}, \\
\hat{P}^{i_1,i_2} \equiv \, & ^{i_1} \leftrightarrow {}^{i_2}, \\
\label{eq:P^i3^i4}
\hat{P}^{i_3,i_4} \equiv \, & ^{i_3} \leftrightarrow {}^{i_4}, \\
\label{eq:P_i3_i4}
\hat{P}_{i_3,i_4} \equiv \, & _{i_3} \leftrightarrow {}_{i_4}, \\
\hat{P}_{a_1,a_3} \equiv \, & _{a_1} \leftrightarrow {}_{a_3}, \\
\hat{P}_{a_2,a_4} \equiv& \, _{a_2} \leftrightarrow {}_{a_4},
\end{align}
as well as their products (e.g., $\hat{P}^{a_3i_1,a_4i_2} \equiv \hat{P}^{a_3,a_4}\hat{P}^{i_1,i_2}$) leave this TN invariant.
    \item {\bf Multiple Identical Tensors}. Appearance of multiple identical tensors in the same tensor network. For example, elements of the tensor $t$ appear twice in \cref{eq:tn1}. Thus, wholesale swapping of the corresponding slots of the two tensor elements leaves the tensor network invariant, that
is, $\hat{P}_{a_1,a_2} \hat{P}_{a_3,a_4} \hat{P}^{i_1,i_3} \hat{P}^{i_2,i_4}$.
    \item {\bf Dummy Indices}. The relabeling of dummy indices can also be viewed as slot permutation, e.g., relabelings $i_3 \leftrightarrow i_4$ and $a_3 \leftrightarrow a_4$ correspond to slot permutations $\hat{P}_{i_3,i_4} \hat{P}^{i_3,i_4}$ and $\hat{P}_{a_3,a_4} \hat{P}^{a_3,a_4}$, respectively. Note that the former can be obtained as a product of two transpositions that we already encountered, namely \cref{eq:P^i3^i4} and \cref{eq:P_i3_i4}, but not the latter.
\end{enumerate}
The unique subset of these slot permutations are {\em generators} of the automorphism group of TN. All possible products thereof, along with the {\em trivial} slot permutation (identity), form the automorphism group of the tensor network. The automorphism group ${\rm Aut}({\bf T})$ of the tensor network ${\bf T}$ with $n$ slots is a subgroup of the symmetric group $S(n)$ that consists of {\em all} possible permutations of $n$ slots. The slot symmetries of a particular TN are described not by the automorphism group itself but by its particular {\em representation}.

The algorithm $f: {\bf T} \to {\bf T}'$ {\em canonicalizes} TN ${\bf T}$ if (a) the permutation of the input ${\bf T}$ by any member of its automorphism group leaves the result unchanged modulo a phase change or complex conjugation (denoted by $\cong$),
\begin{align}
\label{eq:def-canonicalizer-a}
\forall \hat{g}\in \mathrm{Aut}(\mathbf{T}): \quad f(\hat{g}\mathbf{T}) \cong f(\mathbf{T}),
\end{align}
where $\cong$ denotes the equivalence of tensor networks 
and (b) the result is an automorphic permutation of ${\bf T}$,
\begin{align}
\label{eq:def-canonicalizer-b}
  \exists \hat{g}\in \mathrm{Aut}(\mathbf{T}): \quad f(\mathbf{T}) \cong \hat{g}\mathbf{T}.
\end{align}
For general input ${\bf T}$, the cost of computing the canonical form was thought to be in the worst case proportional to the size of Aut({\bf T}), i.e., growing factorially with the number of slots in {\bf T}.\cite{VRG:babai:1983:PFAASTC-S8} However, recently quasipolynomial algorithms have been found.\cite{VRG:babai:2019:P5AASSTC}

The most straightforward (naive) way to canonicalize tensors is through exhaustive search. Namely, for a fixed canonical order of slots in {\bf T} we seek $\hat{g}\mathbf{T}, \hat{g}\in\mathrm{Aut}(\mathbf{T})$ that corresponds to the lexicographically smallest list of indices obtained from the canonical list of slots. Such an approach was used, for example, by Janssen and Schaefer (their search took into account equivalences of TN factors to speed-up the process).\cite{VRG:janssen:1991:TCA}
Exhaustive canonicalization is simple to implement and guaranteed to work for arbitrary {\bf T}. Its cost is proportional to the order of $\mathrm{Aut}(\mathbf{T})$, therefore, can be quite high for large TNs composed of tensors with index symmetries. However, for TNs with particular structure it is possible to design more efficient algorithms. The trivial case is a TN composed of $N$ unique tensors (i.e., every TN factor is distinct); to canonicalize such a TN it is sufficient to sort the TN factors (e.g. lexicographically) in $\bigO{N\log N}$ steps, then canonicalize the indices of each TN factor (typically, also by sorting). Unfortunately, most of the time the TNs one encounters in classical or quantum field-theoretic simulation will  include many identical tensors. For example, consider a variant of TN in \cref{eq:tn1} which includes an explicit antisymmetrizer for the external indices represented as a tensor:\footnote{Operator $\hat{\mathcal{A}}(i_{1},i_{2};a_{1},a_{2})$ antisymmetrizes expression with respect to relabelings $i_1 \leftrightarrow i_2$ and $a_1 \leftrightarrow a_2$, namely,
\begin{align}
 \hat{\mathcal{A}}(i_{1},i_{2};a_{1},a_{2}) f^{i_1 i_2}_{a_1 a_2} \equiv \frac{1}{4} \left( f^{i_1 i_2}_{a_1 a_2} - f^{i_2 i_1}_{a_1 a_2} - f^{i_1 i_2}_{a_2 a_1} + f^{i_2 i_1}_{a_2 a_1}\right).
\end{align}
In tensor expressions the action of such an \emph{operator} $\hat{\mathcal{A}}(i_{1},i_{2};a_{1},a_{2})$ is conveniently expressed as a contraction with \emph{tensor} $\hat{A}^{a_{1} a_{2}}_{i_{1} i_{2}}$ with antisymmetric bra and ket, $\hat{A}^{a_{1} a_{2}}_{i_{1} i_{2}} f^{i_1 i_2}_{a_1 a_2}$,
with implied Einstein summation. Such representation makes it possible to treat expressions with index symmetrizers and antisymmetrizers as ordinary tensor networks for the purposes of symbolic manipulation and, specifically, canonicalization. More details will be given in the follow-up manuscript where such operators will play a major role.}
\begin{align}
\label{eq:tn1-A}
\hat{A}^{a_{1} a_{2}}_{i_{1} i_{2}} \bar{g}^{{a_3}{a_4}}_{{i_3}{i_4}} t^{{i_1}{i_2}}_{{a_1}{a_3}} t^{{i_3}{i_4}}_{{a_2}{a_4}}.
\end{align}
This TN has 2 instances of the same $t$ tensor, and hence wholesale swapping of the slots of the $t$ tensors is in the automorphism group. One could decide the canonical order of the two equivalent tensors $t$ by comparing them lexicographically. For example, if we define the canonical index string of $t^{k_1 k_2}_{b_1 b_2}$ as $\{b_1, b_2, k_1, k_2\}$, then the canonical index strings of the $t$ tensors in \cref{eq:tn1-A} are $\{a_1,a_3,i_1,i_2\}$ and $\{a_2,a_4,i_3,i_4\}$. The first is lexicographically less than the second (assuming $x_i < x_j \leftrightarrow i < j$), so we conclude that the $t$ tensors are already in lexicographic order. Unfortunately, lexicographic ordering is not invariant with respect to other elements of the automorphism group, such as dummy relabeling $a_1 \leftrightarrow a_2$. Domain-specific networks can still be handled via lexicographic ordering; for example, the structure of TNs in the traditional CC can be exploited to define the lexicographic order of $t$ tensors of the same rank, for example, in the context of factorized evaluation of CC equations\cite{VRG:kallay:2001:JCP} and in the context of canonicalization of CC equations produced algebraically\cite{VRG:evangelista:2022:JCP}. However, generic canonicalization cannot rely on such ad hoc heuristics.

The traditional generic approaches to tensor canonicalization are group-theoretic in nature. Butler-Portugal\cite{VRG:butler:1991:,VRG:portugal:1998:CPC,VRG:portugal:1999:JPMG,VRG:manssur:2002:IJMPC,VRG:martin-garcia:2008:CPC}
is the best-known algorithm.
More elaborate approaches also handle nonabelian symmetries.\cite{VRG:tichai:2020:EPJA}

{\em Graph-theoretical} approaches to tensor canonicalization and/or isomorphism have been explored relatively recently. The oldest was the MSc thesis by Obeid\cite{VRG:obeid:2001:}. Recently, such graph-based techniques have been applied in field-theoretic contexts;\cite{VRG:bolotin:2013:,VRG:li:2017:P2AISSAC,VRG:peeters:2018:J,VRG:kryukov:2019:JPCS} hybrids between group-theoretic and graph-theoretic approaches have also been considered.\cite{VRG:niehoff:2018:CPC,VRG:kryukov:2019:JPCS}

Here we designed a graph-theoretic approach to canonicalization that is fast, respects various TN symmetries, can handle noncovariant TNs with hyperedges (e.g., contractions over more than 2 indices) and index dependencies (\cref{sec:tensor:index_dependencies}).
The key idea is to map a TN to a {\em colored} graph in a way that encodes the structure and symmetries of TN. The automorphism group and the canonical order of the colored graph can then be computed
using existing efficient approaches\cite{VRG:mckay:1981:CN,VRG:junttila:2007:PMAEE,VRG:mckay:2014:JoSC}.

\subsubsection{TN Canonicalizer Algorithm}

\begin{figure*}
\begin{tikzpicture}
\node[inner sep=0pt] (formula) at (-2,3)
    {$\boxed{\atensor{B}{*^{i_1}_{a_1^{i_1 i_2}}}[p_1] \, \atensor{B}{*^{i_2}_{a_2^{i_1 i_2}}}[p_1] \, \atensor{\tilde{a}}{*^{a_1^{i_1 i_2}}_{i_1}*^{a_2^{i_1 i_2}}_{i_2}}}$};
\node[inner sep=0pt] (graph) at (2,-1)
    {\includegraphics[width=.75\textwidth]{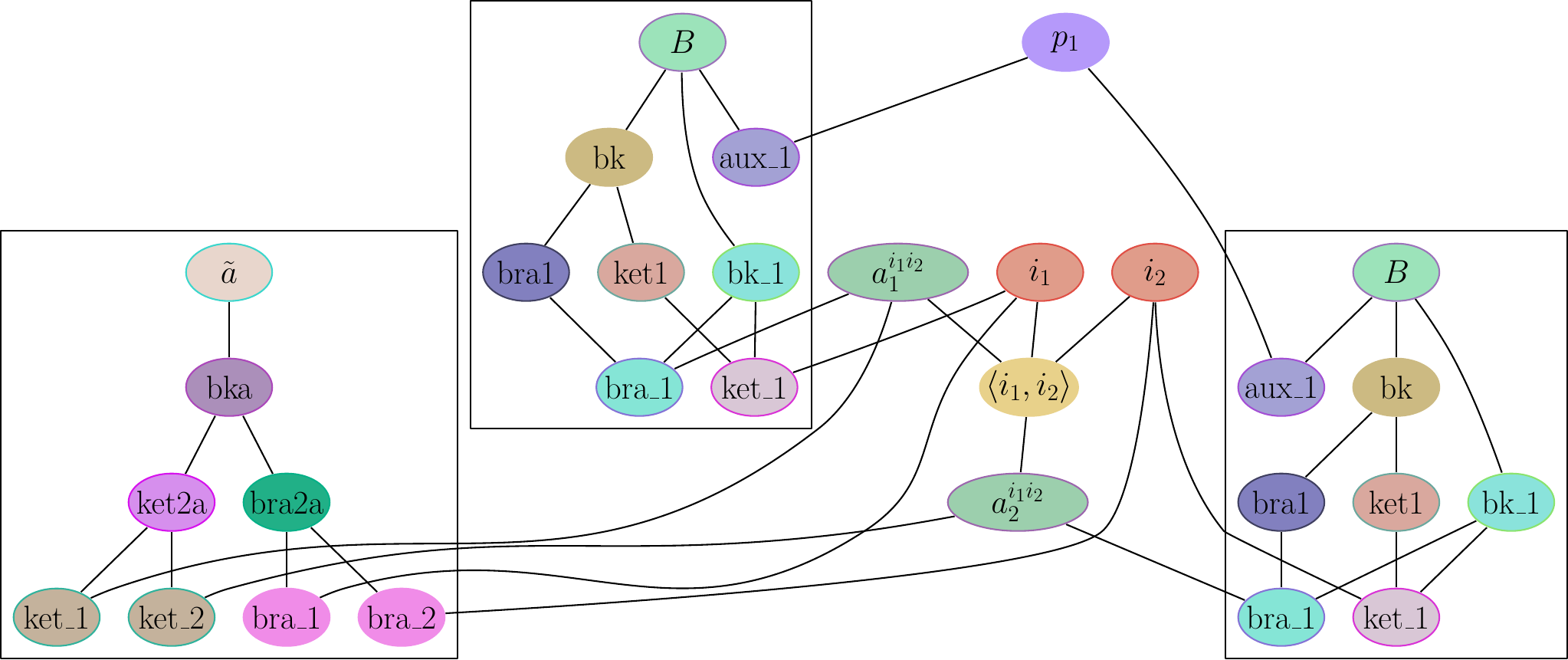}};
\end{tikzpicture}
\caption{An illustrative example of the colored graph representation of a tensor network containing tensors with antisymmetric bra/ket ($\tilde{a}$), tensor with aux indices ($B$), and index dependencies ($a_{k}^{i_1 i_2}$).}
\label{fig:colored-graph-example}
\end{figure*}

Our approach can be best understood by considering a simple example illustrated in \cref{fig:colored-graph-example}. All elements of TN that are needed to specify its identity and symmetries are represented by the graph vertices. That is, all indices, slots, and slot bundles are represented by vertices. Tensor ``cores" (slots for tensor labels) are also represented by vertices. Vertex colors encode the identity of the corresponding objects; two equivalent objects, such as the indices $i_1$ and $i_2$, have identical colors. Graph edges encode relationships between the slots and/or indices, such as occupancy of a slot by an index, composition of a slot bundle out of individual slots or bundles, etc. For example, the index $i_1$ occupies the first bra slot of the tensor $\tilde{a}$ and the ket slot of one of the $B$ tensors, and hence the vertex of the latter is connected by edges to the corresponding vertices of the latter. Note the lack of ability to specify the slot in {\em which} the tensor $B$ in the graph is occupied by the index $i_1$. This emphasizes the power of the graph representation: it is free of the arbitrary choice of which $B$ tensor element is first listed in the mathematical formula. Another key feature of the graph representation is the separation of indices from slots and slot bundles. In this way, it becomes apparent that the tensor elements $B$ in the TN refer to the same tensor as the corresponding subgraphs (identified in \cref{fig:colored-graph-example} by rectangles) are identical in structure (vertices, colors, and edges). Whether the two tensor elements are equivalent in the TN of course depends on the rest of the graph, and will be decided by computing its automorphism.

Now, let us discuss the specifics.
To map a tensor network to a colored graph, we assume that a coloring function \code{color} is available that can map an arbitrary object $K$ to a {\em unique} color $c$, i.e.,
\begin{align}
\forall K_1 \neq K_2: \mathtt{color}(K_1) \neq \mathtt{color}(K_2).
\end{align}
Another function \code{ccolor} must also be available to color an object by combining it with a color of another, again producing a unique color:
\begin{align}
\forall K_1 \neq K_2: \code{ccolor}(K_1, c) \neq \code{ccolor}(K_2, c) \\
\forall c_1 \neq c_2: \code{ccolor}(K, c_1) \neq \code{ccolor}(K, c_2)
\end{align}
Together \code{color} and \code{ccolor} can be used to color any sequence of objects. A variety of known hash functions and hash combining functions can be used for \code{color} and \code{ccolor}, respectively.\cite{VRG:knuth:1998:}
Although different choices of \code{color} and \code{ccolor} will produce different canonical representations, this is sufficient for the purposes of deciding the TN isomorphism.

The algorithm for mapping a TN to a colored graph proceeds as follows.
\begin{enumerate}
 \item Each tensor element produces a tensor {\em core} vertex with the color obtained from the tensor label. The tensor core is used to shade the colors of the slots and slot bundles associated with that tensor element, with their colors computed from the color of the tensor code via \code{ccolor}.
 \item Each tensor's braket bundle generates a vertex connected by an edge to the corresponding core vertex.
 \item Each tensor's bra, ket, and aux bundles produce vertices. The bra and ket vertices of tensors that are invariant with respect to the bra-ket interchange (\code{BraKetSymmetry::symm}) have identical colors. The bra and ket vertices are connected to the braket vertex, whereas the aux vertex is connected to the tensor core vertex directly.
 \item The column bundles of each tensor with asymmetric bra/ket bundles (\code{Symmetry::nonsymm}) produce vertices connected to the tensor core vertex. If the tensor is invariant with respect to a permutation of column bundles (\code{ColumnSymmetry::symm}), every column vertex has the same color, else its color is determined by its ordinal in the sequence of column bundles.
 \item Every tensor index slot produces a vertex. Its color is computed from the type of its bundle (bra, ket, aux) and its ordinal in the bundle. It is connected to the vertex of its bundle. The bra and ket slots of the tensors with asymmetric bra/ket bundles are also connected to the corresponding column vertex.
 \item Each nonnull \code{Index} produces a vertex colored by its \code{IndexSpace}. The vertex is connected to the vertices of the tensor slots occupied by that index in the TN. Each \code{Index} can be connected to any number of aux slots, but at most one bra or one ket slot, thereby tensor networks with hyperedges are nearly fully supported. The restriction against connecting multiple bra or multiple ket slots is for checking against accidental covariance violations in tensor networks; aux slots should be used explicitly to construct noncovariant tensor networks.
 \item Each protoindex bundle produces a vertex colored by the \code{IndexSpace} objects of the constituent indices. The vertex is attached to the vertices of the constituent indices, as well as the vertices of the indices that depend on this index bundle.
\end{enumerate}

Once the colored graph representation of the TN is constructed, the canonical order of its vertices can be computed using one of the well-established approaches. Computing the canonical order of graph vertices is a long-studied problem\cite{VRG:mckay:1978:CM,VRG:babai:1983:PFAASTC-S8,VRG:toran:2004:SJC}
that typically is solved by finding a compact set of generators for the graph automorphism group\cite{VRG:mckay:1978:CM}; both problems are closely related to the graph isomorphism problem.\cite{VRG:whitney:1932:AJM,VRG:lubiw:1981:SJC,VRG:mckay:1981:CN,VRG:mckay:2008:EoA,VRG:mckay:2014:JoSC}
For {\em colored} graphs there are established tools for computing the canonical order and the automorphism group of colored graphs rooted in backtracking algorithm design pioneered by McKay\cite{VRG:mckay:1976:,VRG:mckay:1978:CM,VRG:mckay:1981:CN}; namely, \code{nauty}\cite{VRG:mckay:1978:CM,VRG:mckay:1981:CN}, \code{bliss}\cite{VRG:junttila:2007:PMAEE,VRG:junttila:2011:TaPoAiS}, and others.\cite{VRG:mckay:2014:JoSC}
For historical reasons \code{SeQuant} uses \code{bliss}, but another graph canonicalization solver like \code{nauty} could be used.

Due to space constraints, it is not realistic to describe how such tools work; reader should refer to the recent review of the state of the field\cite{VRG:grohe:2020:CA} and to the original references.
The computed canonical order of the indices can be used to define the canonical order of equivalent tensors, slots, and slot bundles.
\begin{enumerate}
\item Tensors are resorted to according to their ordinals in canonical graph order. Because tensors of operators may not commute, tensor sorting must respect the commutativity of tensors. Thus, bubble sort is used for sorting\footnote{Standard (more efficient) sorting algorithms expect comparisons to obey a strict weak ordering, which implies transitivity of the comparison. However, non-commutativity breaks transitivity: if $a < c$ and $c < b$ are true, this does not necessarily imply $a < b$ in the case where $[a,b] \neq 0$ if '$<$' is the commutativity-respecting comparator, which has to yield false for $a<b$ and $b<a$ even if $a \neq b$ to prevent changing their order. This is known as (non-strict) weak ordering, and is sufficient to implement the bubble sort.}, with each swap of neighboring tensors checking first if the two tensors commute; tensors of scalars always commute, but commutativity of tensor of operators depends on whether nonzero Wick's theorem contractions exist between the two tensors (see \cref{sec:wickengine}). 
\item For each tensor with nontrivial symmetries, slots and slot bundles are reordered according to the canonical graph order.
  \begin{enumerate}
  \item Column bundles are canonically reordered in tensors with asymmetric bra/ket but symmetric with respect to permutation of column bundles.
  \item Braket bundles are canonically reordered in tensors that are symmetric with respect to braket exchange.
  \item The slots in symmetric and antisymmetric bras/kets are canonically reordered; for antisymmetric bras, the phase factor resulting from such permutation is accumulated.
  \end{enumerate}
\end{enumerate}
Lastly, dummy indices are regenerated in the order of appearance in the canonicalized TN.

\begin{figure*}
    \centering
    \includegraphics[height=0.75\textheight]{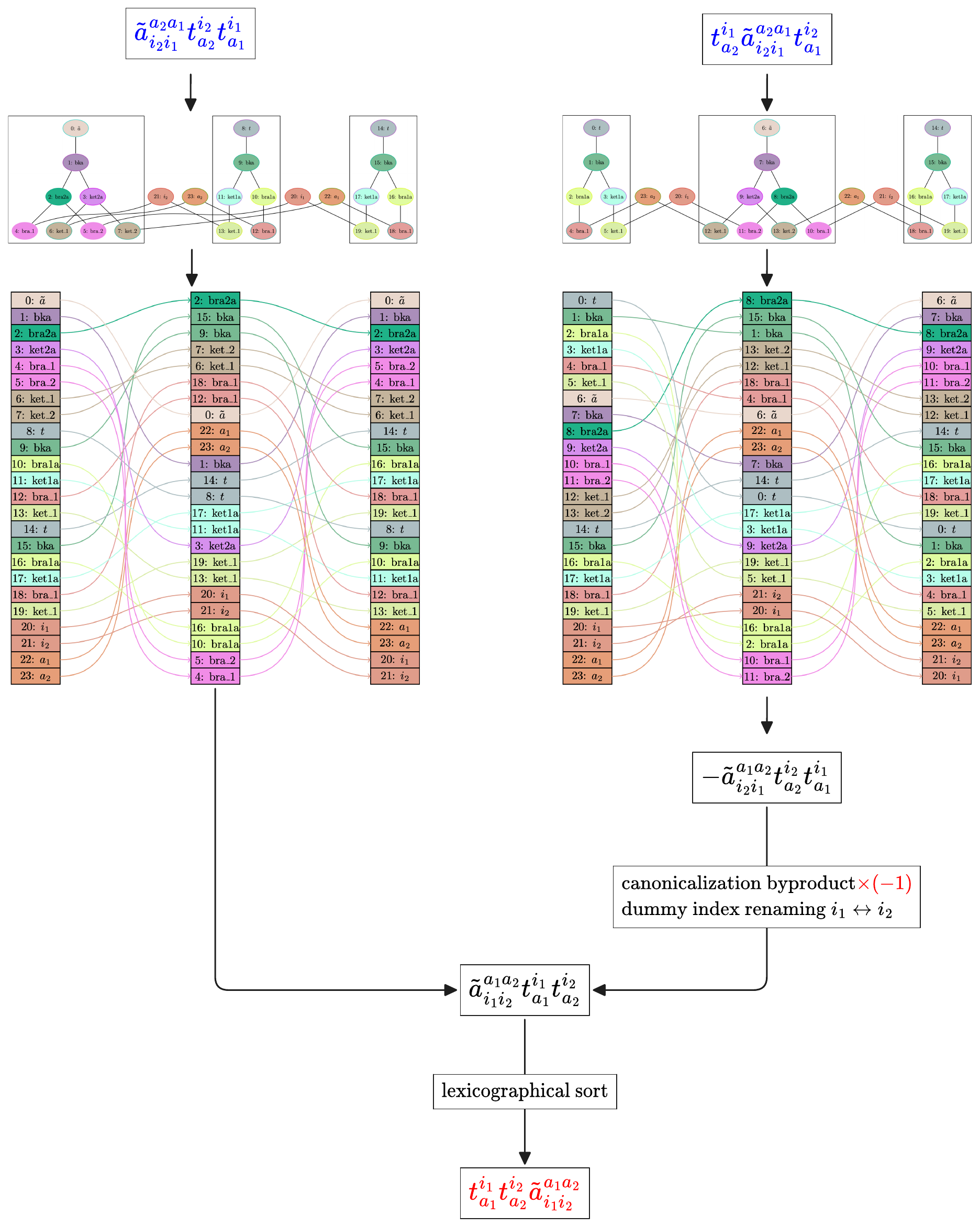}
        \hfill
    \caption{Illustration of TN canonicalization for two simple TNs, $\bm{1}$ (${{\atensor{\tilde{a}}{*^{a_2}_{i_2}*^{a_1}_{i_1}}}{\atensor{t}{*^{i_2}_{a_2}}}{\atensor{t}{*^{i_1}_{a_1}}}}$) and $\bm{2}$ (${{\atensor{t}{*^{i_1}_{a_2}}}{\atensor{\tilde{a}}{*^{a_2}_{i_2}*^{a_1}_{i_1}}}{\atensor{t}{*^{i_2}_{a_1}}}}$), that are equivalent modulo sign. For each TN the corresponding colored graph, its original and canonical order of the vertices, and the regrouping of vertices to match the corresponding tensors are shown. The net permutation of slots of the $\tilde{a}$ tensor in TN $\bm{2}$ produces the sign change. The subsequent renaming of the dummy indices to match to the order of appearance produces identical (modulo sign) tensor network. Lastly, the lexicographic sort of tensors, slots, and slot bundles (with key features like the target order of tensor labels controlled by the user) is designed to produce an aesthetically pleasing final representation; in this case the tensor operators are moved to the right of  tensors of scalars.}
    \label{fig:canonicalization-mechanics}
\end{figure*}

\Cref{fig:canonicalization-mechanics} illustrates the procedure for two sufficiently compact TNs that reduce to the same canonical form modulo sign. Note that the colored-graph-based canonicalization is followed in practice by a rapid lexicographic sort of tensors, slots, and slot bundles designed to produce an aesthetically pleasing representation. If the expressions produced by the canonicalizer are not meant for reading by humans, this step can be skipped. Invoking the \code{canonicalize} function with default options does not invoke lexicographic canonicalization, but the more elaborate \code{simplify} intended for most end-users does.

\subsubsection{TN Canonicalizer Performance}

\begin{figure*}[ht!]
    \centering
    \begin{subfigure}{0.75\linewidth}
    \centering
    \includegraphics[width=\columnwidth]{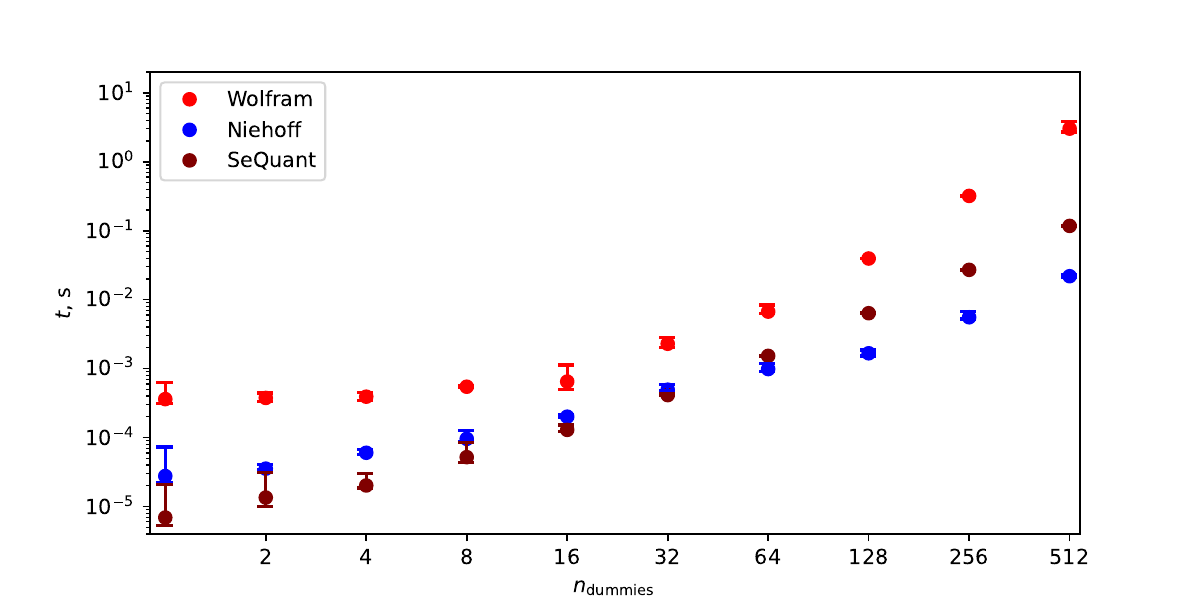}
        \caption{}
        \label{fig:canonicalize-nosym}
    \end{subfigure}\hfill
    \begin{subfigure}{0.75\linewidth}
    \centering
    \includegraphics[width=\columnwidth]{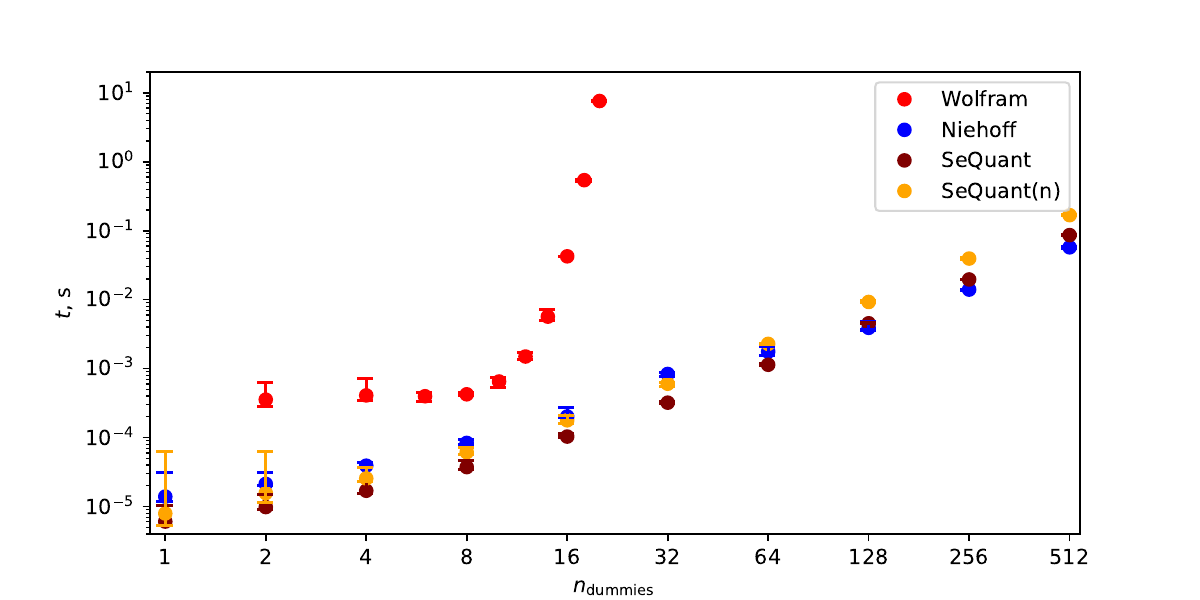}
        \caption{}
        \label{fig:canonicalize-total}
    \end{subfigure}\hfill
    \begin{subfigure}{0.75\linewidth}
    \centering
    \includegraphics[width=\linewidth]{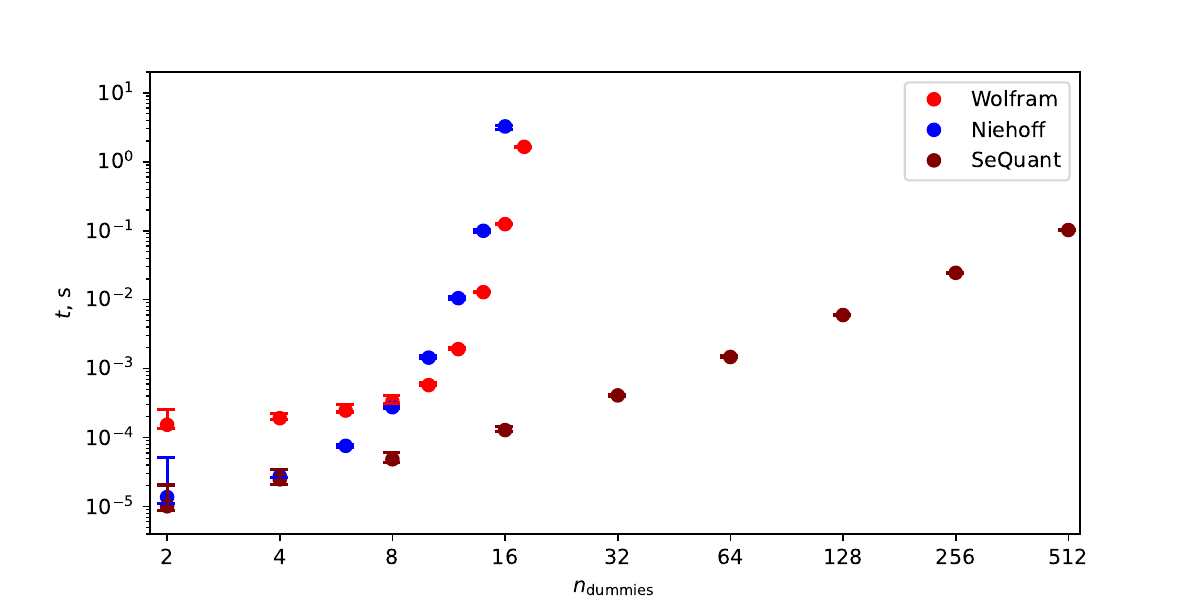}
        \caption{}
        \label{fig:canonicalize-pairwise}
    \end{subfigure}\hfill
    \caption{Comparison of representative tensor canonicalizers (Wolfram: default Butler-Portugal canonicalizer in \code{Mathematica}, Niehoff: the improved Butler-Portugal canonicalizer described in Ref. \citenum{VRG:niehoff:2018:CPC}), SeQuant: the default canonicalizer in \code{SeQuant}. (a) TN in \cref{eq:tensor-total-symm} with non-symmetric tensor $D$; (b) TN in \cref{eq:tensor-total-symm} with totally-symmetric tensor $D$. (c) TN in \cref{eq:tensor-pairwise-symm}. ``SeQuant(n)'' in (b) denotes variant of TN \cref{eq:tensor-total-symm} with symmetric $D_{i_1 \dots i_N}$ represented as a product of equivalent order-1 tensors, $D_{i_1} \dots D_{i_N}$. Timings for Wolfram and Niehoff canonicalizers were obtained using the Wolfram code accompanying Ref. \citenum{VRG:niehoff:2018:CPC}. All timings were obtained on an Apple M1 processor.}
    \label{fig:canonicalize}
\end{figure*}

Our TN canonicalizer can handle large TNs consisting of tensors with several paradigmatic kinds of symmetries, as well as TNs with many equivalent tensors.
To illustrate this, we assessed the performance of our graph-theoretic TN canonicalizer against the standard group-theoretic Butler-Portugal (BP) canonicalizer implemented in the \code{Wolfram} code, and Niehoff's graph-theoretic extension of the BP canonicalizer.\cite{VRG:niehoff:2018:CPC}
Due to the differences in the implementation rigorous comparison is impossible, thus the primary focus will be on the asymptotic scaling of these canonicalizers. We used the following 2 types of tensor networks used for algorithm assessment in Ref. \citenum{VRG:niehoff:2018:CPC}:
\begin{align}
\label{eq:tensor-total-symm}
& D_{i_1 \dots i_N} U^{\pi[i_1 \dots i_N]}, \\
\label{eq:tensor-pairwise-symm}
& D_{i_1i_2} D_{i_3i_4} \dots D_{i_{N-1} i_N} U^{\pi[i_1 \dots i_N]},
\end{align}
where $\pi$ indicates a random permutation. The performance of the 3 canonicalizers averaged over a small random sample of permutations $\pi$ is reported in \cref{fig:canonicalize}.
Whereas for TN \cref{eq:tensor-total-symm} with asymmetric tensors $D$ and $U$ all algorithms perform comparably, with symmetric tensor $U$ the cost of the conventional group-theoretic BP canonicalizer increases dramatically with rank; the graph-theoretic extension of the BP canonicalizer and \code{SeQuant} achieve comparable performance in this case. However, for TN \cref{eq:tensor-pairwise-symm} that contains many equivalent tensors, the cost of both BP variants increased rapidly. We were pleased to find that the fully graph-theoretic canonicalizer in \code{SeQuant} achieves a similar performance for the three TN variants, with the cost growing with the number of slots $n$ approximately as $n^2$.
The polynomial performance of our canonicalizer is not unexpected. Although it is known that the individualization-refinement algorithms used by \code{bliss} and its competitors have worst-case exponential complexity\cite{VRG:neuen:2018:P5AASSTC}
for typical graphs the cost is polynomial. Specifically, for graphs with bounded valence ($O(1)$ edges per vertex), the cost of graph canonicalization is known to be polynomial.\cite{VRG:babai:1983:PFAASTC-S8}
Furthermore, quasipolynomial algorithms ($n^{\mathrm{polylog}(n)}$) for {\em general} canonicalization were recently demonstrated\cite{VRG:babai:2019:P5AASSTC}
and can be deployed if needed.
Whether the asymptotic superiority to the group-theoretic BP approach translates to practical advantage remains to be seen.

\subsection{\code{SQ/tensor}: Wick's Theorem Engine}\label{sec:wickengine}

\subsubsection{Definition}\label{sec:wickengine-definitions}

Wick's theorem (WT) is a systematic algebraic approach for rewriting products of normal-ordered Fock space operators as a superposition of individual normal-ordered operators.\cite{VRG:wick:1950:PR} Namely, product of normal-ordered operators $ABC\dots XYZ$
is the normal-ordered product $\{ABC\dots XYZ\}$ plus all {\em unique} single, double, and higher-order {\em contractions} of creators/annihilators between the normal-ordered operators. The contraction rules are simple for genuine and Fermi vacua, where only contractions between pairs of creators/annihilators are considered; for general vacua hypercontractions (which involve groups of $n$ creators and $n$ annihilators, $n>1$) are also needed.\cite{VRG:mukherjee:1997:CPL,VRG:kutzelnigg:1997:JCP,VRG:kong:2010:JCP}
Here is the simplest example that assumes normal order with respect to the genuine vacuum:
\begin{center}
\begin{minipage}{\columnwidth}
\begin{align}
    \label{eq:WT-a1a1}
    a^p_q a^r_s \equiv & \{a^p a_q\} \{a^r a_s\} \nonumber \\
    \overset{\text{WT}}{=} & \{a^p a_q a^r a_s\} + \{a^p \tikzmark{starta} a_q \tikzmark{enda} a^r a_s\} \nonumber \\
    = & \{a^p a^r a_s a_q \} + \delta^r_q \{a^p a_s\} = a^{pr}_{qs} + \delta^r_q a^p_s,
\end{align}
\JoinUp{0.25}{0.75}{0.25}{0.75}{a}
\end{minipage}
\end{center}
where
\begin{center}
\begin{minipage}{\columnwidth}
\begin{align}
    \label{eq:contraction-physvac}
    \{\tikzmark{startb} a_q \tikzmark{endb} a^r\} = \delta_q^r
\end{align}
\JoinUp{0.25}{0.75}{0.25}{0.75}{b}
\end{minipage}
\end{center}
denotes the contraction whose equality to the Kronecker delta assumes biorthonormality of the bra/ket sp states (see the discussion in \cref{sec:tensor:scalars-and-operators}).
From now on we will denote contractions in the tensor notation directly, with \cref{eq:WT-a1a1} becoming:
\begin{center}
\begin{minipage}{\columnwidth}
\begin{align}
a^p_q a^r_s = a^{pr}_{qs} + \tikzmark{startc}a^p_q \,\tikzmark{middlec}\,\tikzmark{endc}a^r_s = a^{pr}_{qs} + \delta^r_q a^p_s
\end{align}
\JoinDownUp{0.65}{-0.4}{0.65}{-0.8}{c}
\end{minipage}
\end{center}

Note that \Cref{eq:WT-a1a1} is valid for both bosons and fermions; the lack of sign rule is due to the choice of a {\em particular} representation of the normal-ordered operators (which is not unique, in general). That is, the normal-ordered operators in the result maximally preserve the original bra-ket index pairings. This fact is sometimes referred to as the generalized Wick's theorem (GWT).\cite{VRG:kutzelnigg:1982:JCP,VRG:kutzelnigg:2003:ECWFiCaPa}
GWT works not only with conventional but also with spin-free normal-ordered operators.
Extension of GWT to higher ranks is straightforward:
\begin{center}
\begin{minipage}{\columnwidth}
\begin{align}
\label{eq:WT-a2a2}
a^{pr}_{qs} a^{tx}_{vy} = & a^{prtx}_{qsvy} + \delta^t_q a^{prx}_{vsy} + \delta^x_q a^{prt}_{ysv} + \delta^t_s a^{prx}_{qvy} + \delta^x_s a^{prt}_{qyv} \nonumber \\
+ & \delta^t_q \delta^x_s a^{pr}_{vy} + \delta^t_s \delta^x_q a^{pr}_{yv}.
\end{align}
\end{minipage}
\end{center}

With the Fermi vacuum, GWT involves sign rules, both bottom-top and top-bottom contractions occur, and the contractions may produce multiple Kronecker deltas. For example, with 
$i$ representing single-particle (sp) states occupied in the Fermi vacuum, $a$ unoccupied sp states, and $p,q$ any sp state, contractions produce
\begin{center}
\begin{minipage}{\columnwidth}
\begin{align}
\label{eq:contraction-fermivac-particle}
    \{\tikzmark{startd} a_p \,\tikzmark{middled}\,\tikzmark{endd} a^q \} = & \delta_p^a \delta_a^q, \\\nonumber\\
\label{eq:contraction-fermivac-hole}
    \{\tikzmark{starte} a^p \,\tikzmark{middlee}\,\tikzmark{ende} a_q \} = & -\delta^p_i \delta^i_q.
\end{align}
\JoinDownUp{0.65}{-0.4}{0.65}{-0.8}{d}
\JoinUpDown{0.65}{0.7}{0.65}{0.4}{e}
\end{minipage}
\end{center}

Wick's theorem should be viewed as an algorithm for canonicalizing products of (tensor) operators. Note that Wick's theorem does not produce a canonical form of the expression, since the normal-ordered operators can be represented in multiple equivalent forms, e.g., $a^{pr}_{qs} = a^{rp}_{sq}$. However, a subsequent generic canonicalization can be done easily.

\subsubsection{Implementation}\label{sec:wickengine-implementation}

\code{SeQuant} implements Wick's theorem for both bosonic operators (in normal order with respect to the genuine vacuum only) and for fermionic operators (in normal order with respect to genuine or Fermi vacua). While for evaluation of vacuum averages of products of normal operators one is interested in complete contractions only, in general evaluation of all partial contractions is also supported. This is needed, for example, to implement vacuum averages over general (multiconfigurational) Fock-space states.

Wick's theorem is naturally expressed as a recursion, rather than an iteration. The basic outline is simple. For each eligible quasiparticle (qp) annihilator\footnote{Any multiple contraction can be generated using multiple sequences of single contractions, since permuting the sequence  leads to the same multiple contraction. If only full contractions are sought these redundancies can be eliminated by contracting only the first found qp annihilator; furthermore, in such a case the first operator {\em must} be a qp annihilator, otherwise full contraction is not possible. If partial contractions are sought then contractions of every qp annihilator must be considered; to eliminate the possibility of generating the same unique multiple contraction  multiple times only qp annihilators to the right of the most recently contracted qp annihilator are considered for contractions.} scan rightward for a qp creator that's part of another normal-ordered operator. Such annihilator/creator pair can be contracted if the overlap of their spaces is nonnull; if so, remove the two operators from the current product, accounting for the optional fermionic phase (-1 if the number of operators between the contracted operators is odd) and append the contraction result (\cref{eq:contraction-physvac,eq:contraction-fermivac-particle,eq:contraction-fermivac-hole}) to the scalar part of the result. If partial contractions are desired, add the result to the running total, else only do so if no operators remain. If all operators are exhausted, undo the effects of the most recent productive contraction and return to the caller. If no productive contractions were found, return to the caller also.

\subsubsection{Optimizations}\label{sec:wickengine-optimizations}

While Wick's theorem is easy to implement, it is often inefficient because it produces a factorial number of equivalent terms that only differ by the slot reorderings and dummy index renamings. The TN canonicalization procedure described in \cref{sec:canon} can be used to combine the equivalent terms, but this does not help to defray the excessive cost due to the naive implementation of WT. There are several causes for the potential inefficiency of the naive WT and known solutions for each.
\begin{itemize}
\item Typically we are not interested in products of individual matrix elements of tensor operators but in products of tensor operators contracted with tensors of scalars. Consider the following counterpart of \cref{eq:WT-a2a2}, with antisymmetric $g$:
\begin{align}
\label{eq:WT-ga2ga2}
\left( g_{pr}^{qs} a^{pr}_{qs} \right) \left( g_{tx}^{vy} a^{tx}_{vy} \right) = & g_{pr}^{qs} g_{tx}^{vy}a^{prtx}_{qsvy} + 4 g_{pr}^{qs} g_{qx}^{vy} a^{prx}_{vsy} \nonumber \\
& + 2 g_{pr}^{qs} g_{qs}^{vy} a^{pr}_{vy}.
\end{align}
The terms 2-5 and 6-7 on the right-hand side of \cref{eq:WT-a2a2} collapsed into single terms in \cref{eq:WT-ga2ga2} due to the freedom to rename the dummy indices in the latter. This freedom is fundamentally due to the fact that the pairs of slots in \cref{eq:WT-a2a2} that were permutationally related (e.g., $t$ and $x$) but distinguishable due to being occupied by distinct named indices became indistinguishable in \cref{eq:WT-ga2ga2} by the virtue of being occupied by dummy indices. Hence, contractions to $t$ and to $x$ would lead to identical results; we can account for this by contracting to $t$ only and scaling the contraction result by the degeneracy factor of 2 to account for the equivalence of $t$ and $x$.
Such a technique was proposed by Xiao et al in the context of Feynman diagram generation, albeit using different terminology \cite{VRG:xiao:2013:CPC}
. For example, consider the degeneracy factors involved in the application of these rules to the contraction of
the two normal-ordered operators in \cref{eq:WT-ga2ga2}.
The $q\leftrightarrow s$ and $t\leftrightarrow x$ slot equivalences mandate that only the contraction between $q$ and $t$ is needed, weighted by the $\times 4$ degeneracy factor:
\begin{center}
\begin{minipage}{\columnwidth}
\begin{align}
\tikzmark{startf}a^{pr}_{qs} \,\tikzmark{middlef}\,\tikzmark{endf}a^{tx}_{vy} \longrightarrow & \times 4
\end{align}
\JoinDownUp{0.65}{-0.4}{0.65}{-0.8}{f}
\end{minipage}
\end{center}
Any other equivalent contraction must be excluded:
\begin{center}
\begin{minipage}{\columnwidth}
\begin{align}
\tikzmark{startg}a^{pr}_{qs} \,\tikzmark{middleg}\,\tikzmark{endg}a^{tx}_{vy} \longrightarrow & \times 0
\end{align}
\JoinDownUp{0.65}{-0.4}{1}{-0.8}{g}
\end{minipage}
\end{center}
The subsequent single contractions involve slots that are no longer equivalent to any remaining slot, and hence no extra degeneracy factor needs to be added. In fact, a factor of $1/2$ must be added to account for the fact that the 2 contractions are equivalent to each other: 
\begin{center}
\begin{minipage}{\columnwidth}
\begin{align}
\tikzmark{starth}\tikzmark{starti}a^{pr}_{qs} \,\tikzmark{middleh}\,\tikzmark{middlei}\,\tikzmark{endh}\tikzmark{endi}a^{tx}_{vy} \longrightarrow & \times 2
\end{align}
\JoinDownUp{0.65}{-0.4}{0.65}{-0.8}{h}
\JoinDownUp{1.0}{-0.6}{1}{-0.8}{i}
\end{minipage}
\end{center}
The degeneracy factor involving $k$ contractions between bundles of $n$ and $m$ equivalent slots is $n! m! / ((n-k)! (m-k)! k!)$.
\item The same idea applies not only to the individual slots but also to whole tensors. For example, the $a$ operators are distinguishable in \cref{eq:WT-a2a2}, but indistinguishable in \cref{eq:WT-ga2ga2}, hence if both appeared in a larger product, {\em first} contraction to either operator would lead to identical results. We can account for this by contracting to the first of the two equivalent operators and multiplying the contraction result by $2$.
This technique was also proposed by Xiao et al.\cite{VRG:xiao:2013:CPC}
As an illustration, consider the following counterpart of \cref{eq:WT-ga2ga2}:
\begin{align}
\label{eq:WT-ga2ga2-2}
\left( g_{pr}^{qs} a^{pr}_{qs} \right) \left( h_{t}^{v} a^{t}_{v} \right) \left( h_{x}^{y} a^{x}_{y} \right)
\end{align}
The equivalence of operators $a^{t}_{v}$ and $a^{x}_{y}$ allows us to consider the contraction of $q$ to $t$ only, and not to $x$; the resulting degeneracy factor of $\times 4$  arises from the $\times 2$ factor due to the equivalence of $q$ and $s$ slots, and from the $\times 2$ factor due to the equivalence of the operators:
\begin{center}
\begin{minipage}{\columnwidth}
\begin{align}
\tikzmark{startj}a^{pr}_{qs} \,\tikzmark{middlej}\,\tikzmark{endj}a^{t}_{v} \, a^{x}_{y} \longrightarrow & \times 4.
\end{align}
\JoinDownUp{0.65}{-0.4}{0.65}{-0.8}{j}
\end{minipage}
\end{center}
Any other equivalent contraction must be excluded, e.g.:
\begin{center}
\begin{minipage}{\columnwidth}
\begin{align}
\tikzmark{startk}a^{pr}_{qs} \,\tikzmark{middlek}\,\tikzmark{endk}a^{t}_{v} \, a^{x}_{y} \longrightarrow & \times 0.
\end{align}
\JoinDownUp{0.65}{-0.4}{1.8}{-0.8}{k}
\end{minipage}
\end{center}
The subsequent contractions involve slots/operators that are no longer equivalent to any remaining slot/operator, hence no extra degeneracy factor needs to be added. As was the case with the slot equivalences only, a factor of $1/2$ must be added to account for the fact that the 2 contractions are equivalent to each other:
\begin{center}
\begin{minipage}{\columnwidth}
\begin{align}
\tikzmark{startl}\tikzmark{startm}a^{pr}_{qs} \,\tikzmark{middlel}\,\tikzmark{middlem}\,\tikzmark{endl}\tikzmark{endm}a^{t}_{v} \, a^{x}_{y} \longrightarrow & \times 2.
\end{align}
\JoinDownUp{0.65}{-0.4}{0.65}{-0.8}{l}
\JoinDownUp{1.0}{-0.6}{1.8}{-0.8}{m}
\end{minipage}
\end{center}
\item When seeking full contractions, it is possible to run the WT recursion into dead ends by exhausting creators/annihilators ``prematurely'' on some normal-ordered operators. In specific contexts, such as when full contractions are sought using Fermi vacuum WT, it is possible to reduce the number of such situations by keeping track of the number of quasiparticles created or annihilated by each normal-ordered operator; such a technique was already employed, for example, by Janssen and Schaefer.\cite{VRG:janssen:1991:TCA}
\item Sometimes it is known that particular pairs of normal-ordered operators must be connected by one of more contractions. Such connectivity constraints can be checked dynamically during the WT recursion to prune the search space. Not applying the connectivity constraints would produce terms that would need to be filtered out after the WT application.
\end{itemize}

\code{SeQuant}'s Wick's theorem engine implements several of the above techniques to attain high efficiency in a purely algebraic framework.
\begin{itemize}
    \item {\em Topological optimizations}:
    The engine reduces the number of equivalent
    terms produced by the WT recursion by avoiding the equivalent contractions. The equivalence of individual creators/annihilators in slot bundles as well as the equivalence of normal-ordered operators are exploited.
    As already mentioned, Xiao et al proposed exploitation of such techniques in the context of Feyman diagram generation.\cite{VRG:xiao:2013:CPC} However, the Feynman diagram generation is a very narrow use case focused on computing the vacuum averages; hence, all indices are dummy, and all operators (``fields'') of a given type are equivalent. In a general setting it is necessary to boost these techniques to handle named indices (slots occupied by named indices are distinguishable from all others in a given slot bundle) and more general types of TNs involving tensor operators. Using the graph-theoretic methods developed for the purpose of the TN canonicalization in \cref{sec:canon}, it is possible to robustly detect groups of topologically equivalent slots of normal-ordered operators and topologically equivalent normal-ordered operators themselves for a given TN. This information is obtained from the generators of the automorphism group of the colored graph used to represent the given TN.
    \item {\em Connectivity constraints:} The user can inform the engine about the need for contractions between specific normal operators, which is also used to reduce the number of contractions considered.
    \item {\em Coarse-grained concurrency}: Wick's theorem can be efficiently applied to sums of operator products (which is more common in practice than application to a single product), with each individual operator product handled in its own thread.
\end{itemize}

To illustrate the impact of these optimizations, \cref{tab:cc-perf-optimizations} reports the computational expense of deriving the so-called $t$-amplitude equations of traditional truncated coupled-cluster (CC) methods \cite{VRG:coester:1958:NP,VRG:cizek:1966:JCP,VRG:bartlett:2007:RMP} (the equations were validated by numerical evaluation against the reference CC H$_2$O/DZP energies up to excitation rank 5 in Ref. \citenum{VRG:kallay:2001:JCP}). The largest performance improvement stems from the topological optimization; e.g., more than 3 orders of magnitude slowdown is observed for CCSDTQ if the topological optimizations are disabled. The benefit of topological optimizations grows steeply with rank, as expected. The effect of other optimizations is smaller, but still important. The overall performance of the WT engine is excellent, with the coupled-cluster equations up to rank 8 derived in less than a second on a laptop. The performance of \code{SeQuant}'s WT engine for this task is estimated to be one to two orders of magnitude better than the performance of C++-based WT implementation in \code{Wick\&d} (see Table II in Ref. \citenum{VRG:evangelista:2022:JCP}).

\begin{figure}[ht!]
    \centering
    \includegraphics[width=\linewidth]{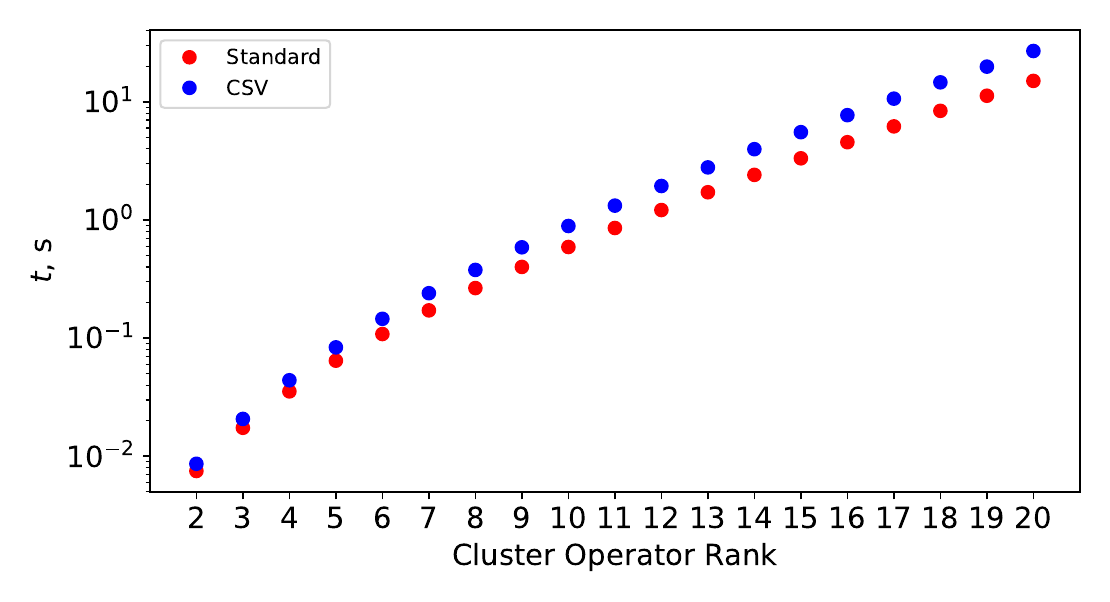}
    \caption{Wall time required to derive the $t$ cluster amplitudes of the standard and cluster-specific (PNO-like) truncated coupled-cluster models. All results were obtained on an Apple M2 processor using 4 threads.}
    \label{fig:cc-performance}
\end{figure}

The optimizations of the WT engine discussed above enable efficient derivation of coupled-cluster amplitude equations in the standard and PNO-like variants (which we refer to here as cluster-specific virtuals) of CC for essentially unlimited ranks (\cref{fig:cc-performance}). The computational expense of deriving the CC equations scales polynomially with the rank of the cluster operator despite the formally factorial number of possible contractions, demonstrating the efficiency of the symbolic engine. The higher computational expense of deriving PNO-style equations is only slightly higher than that of standard equations, since the former expressions will always involve more indices than the latter. 

\begin{table*}[!ht]
    \caption{Impact of the Wick's Theorem engine optimizations (\cref{sec:wickengine-optimizations}) on the computational cost of deriving the traditional ground-state coupled-cluster methods.}
    \label{tab:cc-perf-optimizations}
    \begin{tabular}{rrrrr}
    \toprule
    Method & Wall time (s)$^a$ & Topology$^b$ & Connectivity$^c$ & Threads$^d$ \\
    \midrule
    CCSD & ${7.55 \times 10^{-3}}$ & {5.87} & {3.70} & {1.41}\\
    CCSDT & ${1.75 \times 10^{-2}}$ & {88.5} & {6.54} & {1.61}\\
    CCSDTQ & ${3.55 \times 10^{-2}}$ & {2630} & {10.1} & {1.73}\\
    CCSDTQP & ${6.43 \times 10^{-2}}$ & {} & {13.5} & {1.78}\\
    CCSDTQPH & ${1.08 \times 10^{-1}}$ & {} & {17.6} & {1.86}\\
    CCSDTQPH7 & ${1.75 \times 10^{-1}}$ & {} & {21.8} & {1.93} \\
    CCSDTQPH78 & ${2.68 \times 10^{-1}}$ & {} & {26.4} & {1.99} \\
    \bottomrule
    \end{tabular}
    
    $^a$ Wall times obtained with all optimizations enabled on an Apple M2 processor using 4 threads. \\
    $^b$ Slowdown due to disabling topological optimizations. \\
    $^c$ Slowdown due to disabling connectivity constraints. \\
    $^d$ Slowdown due to reducing the number of threads to 1. \\
\end{table*}

\subsubsection{Wick's Theorem vs Diagrams}\label{sec:wickengine-vs-diagrams}

Let us comment on the relationship between the topological optimization of Wick's theorem introduced here to the diagrammatic techniques\cite{VRG:feynman:1949:PR,VRG:goldstone:1957:PRSMPES,VRG:hugenholtz:1957:P,VRG:brandow:1967:RMP} which are widely employed for derivation and interpretation of the unique terms in products of Fock-space operators, such as those arising from series expansion of time propagators and partition functions in field-theoretic context as well as in many-body methods of more ``traditional'' quantum mechanics. Diagrams also are useful in the context of designing efficient computational strategies for the tensor algebra.\cite{VRG:kucharski:1986:AiQC,VRG:kucharski:1991:TCA} Although some statements in the literature\cite{VRG:bochevarov:2004:JCP,VRG:shiozaki:2008:PCCP} and the sentiment among some segment of the electronic structure community may be interpreted as Wick's theorem being less efficient than the diagrammatic approaches, since the naive application of the former will generate many (factorially many) equivalent terms that need to be canonicalized efficiently, whereas the latter are easily automated to generate only the unique (nonredundant) terms only\cite{VRG:kaldor:1976:JCP,VRG:shiozaki:2008:PCCP,VRG:tag:2017:IJMPC}.

The diagrammatic approaches are nothing but context-specific applications of Wick's theorem that leverage expressions' topological structure (e.g., in traditional coupled-cluster the equivalence of all cluster operators of same rank, symmetry/antisymmetry of the operators) or context-specific structure (e.g., quasiparticle excitation rank, as in ``excitation rank must add up to zero for the vacuum expectation value to be nonzero'') or both. These qualities are both the strength and weakness of the diagrammatic approaches for the derivation purposes: the context-specific notions limit the applicability of the diagrammatic techniques to the particular contexts, whereas the core algebraic rules are applicable universally; particular examples include methods with multideterminantal vacuum\cite{VRG:kutzelnigg:1997:JCP,VRG:kong:2010:JCP} or multicomponent methods.\cite{VRG:bochevarov:2004:MP} The desire for as context-neutral functionality as possible is what motivated our choice of Wick's theorem as the core engine for operator product rewrites. The topological optimizations of the Wick's theorem engine blur the line between the pure diagram-based methods and the naive/inefficient use of the Wick's theorem while maintaining universal applicability. Additional optimizations could be introduced into the WT engine for context-specific cases (such as traditional coupled-cluster), but such optimizations within the WT engine are less important and would make the code more complex and domain-specialized. As will be shown in the follow-up manuscript, such optimizations can be introduced in the higher-level domain-specific abstraction layers.

\subsubsection{Reduction Rules}\label{sec:wickengine-reductionrules}

The application of Wick's theorem in general must be followed by reduction of the Kronecker deltas or equivalents. Such reduction can be fairly complex, especially when non-biorthogonal sp states are involved and noncovariant tensor networks are involved. Here we briefly illustrate the issues; reader is referred to the implementation details for the complete details.

Consider the following operator product defined in the index space context illustrated in \cref{fig:ISR}:
\begin{align}
\tilde{a}_{p_1^{i_1}} \hat{h}_1^\text{d} \tilde{a}^{p_2^{i_2}}
\end{align}
where $p_1^{i_1}, p_2^{i_2}$ index orbital-specific basis (see \cref{sec:tensor:index_dependencies}) spanning the $\{p\}$ space and $\hat{h}_1^\text{d}$ is the diagonal 1-body operator:
\begin{align}
  \hat{h}_1^\text{d} \equiv \sum_p \tilde{a}^p_p \, h[p].
\end{align}
First, consider the contraction to be between $\tilde{a}_{p_1^{i_1}}$ and $\tilde{a}^{p_2^{i_2}}$, and it is nonzero only if $p_1^{i_1}$ and $p_2^{i_2}$ are particle indices (i.e., they are within the $\{a\}$ subspace). Furthermore, when $i_1 \neq i_2$ bases, $\{ p_1^{i_1}\}$ and $\{p_2^{i_2}\}$ are not biorthogonal. Thus, the contraction result involves 2 Kronecker deltas restricting $p_1^{i_1}$ and $p_2^{i_2}$ to the matching (orbital-dependent) bases for the $\{a\}$ subspace and their overlap:
\begin{center}
\begin{minipage}{\columnwidth}
\begin{align}
\tikzmark{startn}\tilde{a}_{p_1^{i_1}} \,\tikzmark{middlen}\,\tikzmark{endn}\tilde{a}^{p_2^{i_2}} = & \delta_{p_1^{i_1}}^{a_1^{i_1}} s_{a_1^{i_1}}^{a_2^{i_2}} \delta^{p_2^{i_2}}_{a_2^{i_2}}
\end{align}
\JoinDownUp{0.65}{-0.4}{0.65}{-0.8}{n}
\end{minipage}
\end{center}
The other two possible contractions are similar:
\begin{center}
\begin{minipage}{\columnwidth}
\begin{align}
\tikzmark{starto}\tilde{a}_{p_1^{i_1}} \,\tikzmark{middleo}\,\tikzmark{endo}\tilde{a}^{p} = & \delta_{p_1^{i_1}}^{a_1^{i_1}} s_{a_1^{i_1}}^{a_1} \delta^{p}_{a_1}
\end{align}
\JoinDownUp{0.65}{-0.4}{0.65}{-0.8}{o}
\begin{align}
\tikzmark{startp}\tilde{a}_{p} \,\tikzmark{middlep}\,\tikzmark{endp}\tilde{a}^{p_2^{i_2}} = & \delta_{p}^{a_1} s_{a_1}^{a_2^{i_2}} \delta^{p_2^{i_2}}_{a_2^{i_2}}
\end{align}
\JoinDownUp{0.65}{-0.4}{0.65}{-0.8}{p}
\end{minipage}
\end{center}

Note that contractions that involve indices with pure quasiparticle character and without index dependencies (which break the biorthogonality) are simpler.

Application of Wick's theorem produces one 0-contraction term, three 1-contraction terms, and one 2-contraction term:
\begin{align}
\tilde{a}_{p_1^{i_1}} \hat{h}_1^\text{d} \tilde{a}^{p_2^{i_2}} \overset{\text{WT}}{=} &
- \sum_p \tilde{a}^{p p_2^{i_2}}_{p p_1^{i_1}} \, h[p] + \sum_p \tilde{a}^{p}_{p} \delta_{p_1^{i_1}}^{a_1^{i_1}} s_{a_1^{i_1}}^{a_2^{i_2}} \delta^{p_2^{i_2}}_{a_2^{i_2}} \, h[p] \nonumber \\
& - \sum_p \tilde{a}^{p_2^{i_2}}_{p} \delta_{p_1^{i_1}}^{a_1^{i_1}} s_{a_1^{i_1}}^{a_2} \delta^{p}_{a_2} \, h[p] \nonumber \\
& - \sum_p \tilde{a}^{p}_{p_1^{i_1}}  \delta_{p}^{a_1} s_{a_1}^{a_2^{i_2}} \delta^{p_2^{i_2}}_{a_2^{i_2}} \, h[p] \nonumber \\
& + \sum_p \delta_{p_1^{i_1}}^{a_1^{i_1}} s_{a_1^{i_1}}^{a_3} \delta^{p}_{a_3} \delta_{p}^{a_4} s_{a_4}^{a_2^{i_2}} \delta^{p_2^{i_2}}_{a_2^{i_2}} \, h[p] 
\label{eq:reduction-example}
\end{align}
To reduce these expressions Kronecker deltas that involve dummy indices are converted to index replacement rules. For example, each Kronecker delta involving 1 dummy and 1 named index adds a replacement rule converting  the dummy index to the named index (if the latter refers to a subspace of the former) or to a new dummy index referring to the intersection of the original indices' spaces (if the former does not refer to a subspace of the latter). Kronecker deltas involving two dummy indices are processed similarly, where Kronecker deltas involving two named indices do not generate replacement rules. Multiple replacement rules involving the same source index, such as replacement rules involving index $p$ in the last term of \cref{eq:reduction-example}, are merged into a single rule to the intersection of their destination spaces, etc. The replacements are applied until the expression stops changing. Finally, trivial simplifications are performed (e.g., Kronecker deltas or overlaps with same bra and ket indices are replaced by 1).

The final result (with ``noncovariant'' summations over $p$ and $a$ implied) reads:

\begin{align}
\tilde{a}_{p_1^{i_1}} \hat{h}_1^\text{d} \tilde{a}^{p_2^{i_2}} \overset{\text{WT}}{=} &
    - {{\tilde{a}^{{p}{p_2^{{i_2}}}}_{{p}{p_1^{{i_1}}}}}\,{h^{}_{}[{p}]}} + {{\tilde{a}^{{p}}_{{p}}}{\delta^{{a_1^{{i_1}}}}_{{p_1^{{i_1}}}}}{s^{{a_2^{{i_2}}}}_{{a_1^{{i_1}}}}}{\delta^{{p_2^{{i_2}}}}_{{a_2^{{i_2}}}}}\,{h^{}_{}[{p}]}} \nonumber \\
    & - {{\tilde{a}^{{a_1}}_{{p_1^{{i_1}}}}}{s^{{a_2^{{i_2}}}}_{{a_1}}}{\delta^{{p_2^{{i_2}}}}_{{a_2^{{i_2}}}}}\,{h^{}_{}[{a_1}]}} - {{\tilde{a}^{{p_2^{{i_2}}}}_{{a_2}}}{\delta^{{a_1^{{i_1}}}}_{{p_1^{{i_1}}}}}{s^{{a_2}}_{{a_1^{{i_1}}}}}\,{h^{}_{}[{a_2}]}} \nonumber \\
    & + {{\delta^{{a_1^{{i_1}}}}_{{p_1^{{i_1}}}}}{s^{{a_3}}_{{a_1^{{i_1}}}}}{s^{{a_2^{{i_2}}}}_{{a_3}}}{\delta^{{p_2^{{i_2}}}}_{{a_2^{{i_2}}}}}\,{h^{}_{}[{a_3}]}}.
    \label{eq:reduction-example-result}
\end{align}
\Cref{listing:reduction-example} shows how \cref{eq:reduction-example-result} can be evaluated and validated in \code{SeQuant}. This example illustrates how \code{SeQuant} expressions can be constructed not as C++ expressions but by parsing their string representations. Such representations can be viewed as a domain-specific language (DSL), with sufficient features for encoding tensors of scalars and operators, index dependencies, and (not illustrated) tensor symmetries. The DSL parsing is performed by the \code{Boost.Spirit} library. Examples of the DSL use can be found throughout the \code{SeQuant} test suite.

\begin{widetext}
\begin{center}
\begin{listing}[H]
\caption{Evaluation and validation of \cref{eq:reduction-example-result} in \code{SeQuant}.}
\label{listing:reduction-example}
\begin{minted}{c++}
auto input = fannx(Index{"p_1", {L"i_1"}}) *
             (ex<Tensor>(L"h", bra{}, ket{}, aux{L"p_3"}, Symmetry::Nonsymm) *
              ex<FNOperator>(cre({L"p_3"}), ann({L"p_3"}))) *
             fcrex(Index{"p_2", {L"i_2"}});
auto result = FWickTheorem{input}.full_contractions(false).compute();
simplify(result);

// expressions can be constructed from LaTeX-like text format
auto expected =
  deserialize("- h{;;p_3} ã{p_1<i_1>,p_3;p_2<i_2>,p_3}"
    "+ h{;;p_3} δ{p_1<i_1>;a_1<i_1>} δ{a_2<i_2>;p_2<i_2>} ã{p_3;p_3} s{a_1<i_1>;a_2<i_2>} "
    "- h{;;a_1} δ{a_2<i_2>;p_2<i_2>} ã{p_1<i_1>;a_1} s{a_1;a_2<i_2>} "
    "- h{;;a_2} δ{p_1<i_1>;a_1<i_1>} s{a_1<i_1>;a_2} ã{a_2;p_2<i_2>} "
    "+ h{;;a_3} δ{p_1<i_1>;a_1<i_1>} δ{a_2<i_2>;p_2<i_2>} s{a_1<i_1>;a_3} s{a_3;a_2<i_2>} ");
simplify(expected);

assert(result == expected)
\end{minted}
\end{listing}
\end{center}
\end{widetext}

\subsection{\code{SQ/tensor}: Index Space Registry}\label{sec:isr}

Many tensor algebra domains involve index spaces equipped with set-theoretic operations (union, intersect) as well as involve assigning index spaces with various traits, such as custom quantum numbers (like spin quantum numbers, identity of particles they represent, etc.) or whether a genuine creator/annihilator in a particular index space is a creator/annihilator with respect to a particular Wick vacuum. \code{SeQuant} allows users to define their own index spaces, their set-theoretic relationships, their traits, and naming conventions.

Class \code{IndexSpace} represents an index space. 
Each index space is a base space or a union of several {\em base} spaces. Thus, the index space is encoded by a 32-bit word whose bits denote its composition in terms of base spaces. Bits of another 32-bit word encode the quantum numbers of the index space (these are primarily used for quantum many-body simulation and will be discussed in the second paper). Class \code{IndexSpaceRegistry} contains the known \code{IndexSpace} objects and the associated metadata, such as the list of base spaces, etc.

\begin{figure}[ht!]
    \centering
    \includegraphics[width=\linewidth]{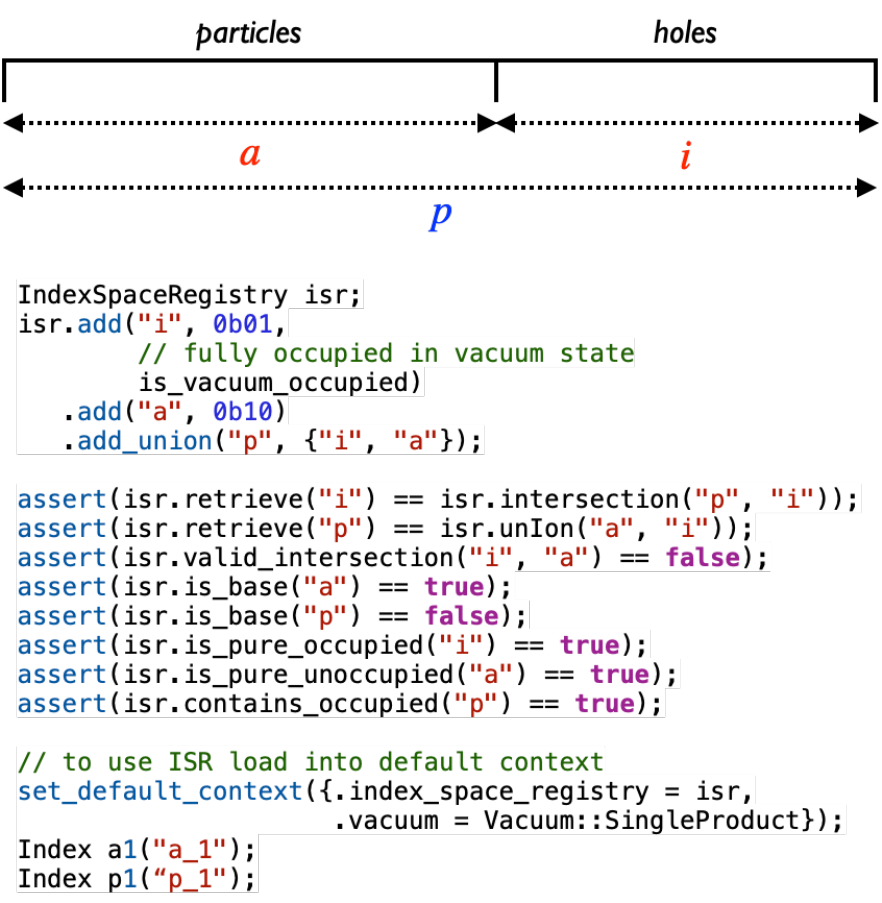}
    \caption{Use of \code{IndexSpaceRegistry} to define a set of two base index spaces.}
    \label{fig:ISR}
\end{figure}

\cref{fig:ISR} illustrates how to define a set of index spaces built from two base spaces with \code{IndexSpaceRegistry}. To make such index space set useful for fermionic many-body context the base spaces are annotated by their roles in Wick's theorem vacuum.
The registry supports set-theoretic operations (computing intersections and unions). Since it is not possible to automate labeling of index spaces, hence it is user's responsibility to declare ahead of time all unions of base spaces that will be encountered (not all possible unions may be needed). The registry supports introspection of spaces (whether a space is a base space, whether it is fully occupied in the Wick vacuum, etc.). 

Once constructed, the index registry can be added to the default context. The custom ``vocabulary'' of index labels defined by the registry can then be used to construct \code{Index} objects and tensors of various types directly from strings.

\section{\code{SQ/eval}: evaluating tensor algebra}\label{sec:eval}

For most users, symbolic manipulation of tensor expressions is only the first step; the ultimate goal is to evaluate the expressions numerically. Efficient evaluation of tensor expressions usually requires significant transformations, such as factorization, common subexpression elimination, etc., with most of these optimizations involving NP-complete subproblems and thus in practice heavily relying on heuristics.
Furthermore, to make rapid method development possible, it is desirable to avoid code generation. For this reason, the primary mode of numerical evaluation of the tensor algebra in \code{SeQuant} is {\em interpretation} at runtime. Since expensive optimizations cannot be used in the interpreter, the interpreter design aims to make its performance comparable --- i.e. not worse by orders of magnitude --- to that of a manually implemented counterpart, rather than try to approach it as closely as possible.

The numerical evaluation sub-module of the \code{SeQuant} library is structured similarly to a modern three-stage compiler, with the representation of the computation gradually lowered from the input \code{SeQuant} expressions to the final optimized representation in terms of basic (unary and binary) operations on scalars and tensors. Unlike a traditional compiler, however, which receives as input a program expressed as text which needs to be parsed and converted to an Abstract Syntax Tree (AST) representation of the input program, the evaluation module receives \code{SeQuant} expressions as input, which directly serve as the AST themselves.

The AST is then lowered to the intermediate representation (IR) which encodes further details of the computation that make it possible to evaluate it (\cref{sec:IR}). Whereas \code{SeQuant} expressions are generally graphs, IR is a simple binary tree because we assume the operation basis available for the evaluation of tensor algebra consists of unary and binary operations (\cref{sec:operation-basis}).
For example, while AST's \code{Product} node represents a scaled product of multiple factors which permits many possible evaluation orders, its IR representation encodes the specific sequence of binary multiplications used to evaluate it as well as other metadata essential for evaluation (such as physical layout of the tensors) and IR manipulation (e.g., unique identifiers for common sub-expression elimination).
IR undergoes multiple passes of rewriting whose goal is to optimize its evaluation.

Finally, the optimized IR is then interpreted directly, by invoking an external numerical tensor framework, or converted to code (transpiled).
In the code generation mode, \code{SeQuant} acts as a transpiler, translating the optimized IR into high-level source code (e.g. C++, Python) that contains calls to the target numerical tensor algebra framework. The generated code is then compiled alongside a host application. This approach allows \code{SeQuant} to serve as a domain-specific front end for lower-level tensor compilers such as TACO \cite{kjolstad2017}, offloading the task of hardware-specific kernel generation while retaining control over high-level domain-aware symbolic optimizations. This capability is under active development and will be the subject of a future publication.

Interpretation has been the primary execution model in \code{SeQuant} because it has significant advantages over code generation, namely:
\begin{itemize}
\item {\bf Cost}: relevant tensor expressions can be quite large (thousands of terms) and the cost of their compilation can be steep, especially with modern high-performance languages. Although one-time compilation costs are moderate, this greatly slows down the development cycle.
\item {\bf Optimization}: interpretation allows deeper optimization by having access to the problem metadata (e.g., dimensions of index spaces, sparsity, etc.), resource information (e.g., amount of memory, number of executor units) which allow more accurate models of performance.
\item {\bf Introspection}: it is easier to debug or profile evaluation of expressions by the interpreter due to being able to instrument the interpreter code (e.g., to print result of every tensor addition operation) directly; with the code generation approach, such changes require code regeneration and rebuilding of the consuming program.
\item {\bf Maintenance}: The lack of a separate code generation step makes it easier to maintain programs that use \code{SeQuant} expressions.
\end{itemize}

In this section, we will primarily focus on the numerical interpretation capabilities of \code{SeQuant}.
First, we will discuss the types of mathematical operations encountered in the tensor algebra supported by \code{SeQuant} (\cref{sec:operation-basis}). We will then discuss the necessary transformations of the expressions into a form that allows an efficient numerical evaluation by the interpreter (\cref{sec:interpretation}).

\subsection{Operation Basis of Tensor Algebra}\label{sec:operation-basis}

Expressions that can serve as input for numerical evaluation can involve scalars (constants \code{Constant} and variables \code{Variable}) and tensors over scalar rings (\code{Tensor}). Numerical evaluation of these ultimately reduces to the basic arithmetic operations on scalars (addition, subtraction, multiplication).
These comprise the entire operation basis for the constants and variables.

The operation basis for tensors is much richer than for scalars. Although we are increasingly facing the need to consider $n$-ary operations as part of the operation basis,\cite{VRG:raje:2024:ATACO} currently \code{SeQuant} is limited to unary and binary operations in the operation basis of scalars. Even then the operation basis of tensors is quite extensive and needs to be discussed in detail. Furthermore, the operation basis further grows when tensors with dependent indices are involved. As discussed in \cref{sec:tensor:index_dependencies} the natural representation of such tensors is by tensors of tensors (TT). The operation basis of TTs is still larger than that of plain tensors.

To classify operations, we denote scalars by S and standard tensors of scalars T. For example, S-T denotes a binary operation that involves a scalar and a tensor of scalars. Similarly, TT-T denotes a binary operation involving a tensor of tensors and a tensor of scalars.

\subsubsection{Tensors of Scalars}
Unlike for scalars, for tensors it is necessary to consider a unary operation that changes its layout, i.e., mode {\em permutation}. This is needed for 2 reasons.
\begin{itemize}
\item To work around the limitations of the numerical backend used to implement the tensor algebra. For example, the \code{BTAS} tensor framework that is used for testing purposes does not support the addition of 2 tensors with different layouts of their modes, and hence explicit permutation of one of the arguments must be performed first. The \code{TiledArray}\cite{VRG:calvin:2015:I15WIAAA} tensor framework used for production purposes does not have this limitation and therefore does not require this operation.
\item To permute the final result of an expression to its user-requested layout. As elaborated below, all intermediate tensors are kept in the layout determined by the TN canonicalizer, which may not be the layout that the user wants. This necessitates permutation to the target layout. 
\end{itemize}
To denote the layout of a tensor in the context of an operation
, we will use index sequences; for example, binary operations involving tensors laid out as $ijk$ and $jki$ producing a tensor laid out as $kji$ will be denoted as $ijk,jki\to kji$.

The only S-T binary operation is the multiplication of a tensor by a scalar value.

T-T binary operations (addition, multiplication)
in general involve 2 operands with arbitrary layouts. For simplicity, we will not distinguish binary operations that differ only by the layout of its arguments and/or the result, even though their efficient implementations may differ. In other words, tensor additions $ijk,ijk \to ijk$, $ijk,ikj\to ijk$, and $jki,ikj\to ijk$ will be considered examples of the same operation type.

The binary T-T product is most complex as there are a number of distinct types of products of tensors (which we will refer in the arithmetic context to as tensor products, not to be confused with other uses of such term) that differ in semantics and computational efficiency. Unlike tensor addition, where the orders of both inputs and the result are all equal, in a product the orders of inputs and the result will in general differ. It is important to first enumerate what can happen to an input tensor mode in a product; for convenience, let's use product $ijkl,jklm\to mijl$.
\begin{itemize}
\item the modes denoted by indices $j$ and $l$ in both inputs and the output are co-iterated over; we will refer to such indices/modes as {\em batching}.
\item the modes denoted by indices $k$ and $k$ in both inputs are summed over, with no matching mode in the output; we will refer to such indices/modes as {\em contracted}.
\item modes denoted by index $i$ in the first input and $m$ in the second input, with the matching modes in the output; we will refer to such indices/modes as {\em free}.
\end{itemize}
The semantics of the $ijkl,jklm\to mijl$ tensor product can be understood by the mathematical expression:
\begin{align}
C_{mijl} = \sum_k A_{ijkl} B_{jklm},
\end{align}
with the summation over the contraction index $k$ shown explicitly. Despite the lack of apparent differences between the batching ($j l$) and free ($i m$) indices in the mathematical expression of the product, there are important differences in their effect on the implementation details and performance; thus, we keep the two types of modes/indices distinct.

As an example that highlights the importance of passing the target indices to encode the type of tensor operation during numerical evaluation, consider the following pseudo-spectral decomposition \cite{VRG:ringnalda:1990:JCP} of two-electron integrals $\left(μ σ|ν λ\right) = \sum_{p} X_{μ p}^{*} Y_{ν λ p} X_{σ p}$. If we strictly adhere to the convention that any repeating index in two tensors is summed over, the right-hand side cannot represent the left-hand side because it becomes an order-5 tensor instead of an order-4 tensor.

\subsubsection{Tensors of Tensors}

The operational basis for tensors of tensors (TT) is conceptually analogous to that of conventional tensors of scalars (T). The key distinction lies in the component-wise operations; whereas for T the fundamental ``ring" operations are scalar addition and multiplication, for TT these are replaced by the tensor operations themselves. We define the component-wise operational basis for TT to be the complete operational basis of T.

To formalize this, we adopt an index notation that separates the indices of the outer tensor from those of the inner tensors with a semicolon. For example, $i j; a b$ denotes an order-2 TT whose components, indexed by $i$ and $j$, are themselves rank-2 Ts with indices $a$ and $b$. This structure allows for a rich set of operations that can be categorized by their action on the outer and inner indices.

\begin{enumerate}
\item \textbf{Unary and Element-wise Operations}

Operations acting on a single TT or element-wise between two TTs are the most
straightforward extensions of conventional tensor algebra.

\begin{itemize}
\item \textbf{Scalar multiplication}: A scalar multiplying a TT scales every inner tensor component uniformly.
\item \textbf{Addition}: The addition of two TTs, $i j; a b + i j; a b \to{} i j; a b$, is defined as the element-wise addition of the corresponding inner tensors. This implies that the inner tensors at each coordinate $(i,j)$ are addition-congruent (i.e., have identical shapes, modulo possible permutation if required by the operation).
\item \textbf{Permutation}: Permutation in a TT is more involved than in a conventional tensor. It can be applied exclusively to the outer indices $(i j k; a b \to{} j i k; a b)$, exclusively to the inner indices ($i j;a b \to{} i j; b a$), or to both simultaneously.
\end{itemize}

\item \textbf{Binary TT Products}

Binary products involving TTs can be classified by whether the contraction occurs on the outer indices, the inner indices, or both.

\begin{itemize}
\item \textbf{Inner contractions only}: When no contraction occurs in the outer indices, the operation is effectively a batched tensor product. For example, in $i j; a b d, i k;a c d \to{} i j k; a b c$, the Hadamard index $i$ batches the operation, while the external indices $j$ and $k$ form an outer product. For each component of the resulting outer tensor, the corresponding ``multiplication" is the inner tensor contraction $a b d, a c d \to{} a b c$.
\item \textbf{Outer contractions}: Contractions over outer indices introduce a crucial constraint. Consider the operation $i k; a c, k j; c b \to{} i j; a b$. The contraction over the outer index $k$ implies a summation of the resulting inner tensors. For this sum to be well-defined, the inner tensors produced by the component-wise multiplications must be addition-congruent. This means that the dimensions of the non-contracted inner indices ($a$ and $b$ in this case) must be uniform across the entire range of the contracted outer index $k$.
\item \textbf{Mixed-rank products (TT $\times{}$ T $\to{}$ TT)}: Operations involving a TT and a conventional tensor T are also common. In a product such as $i, j; a b \to{} i j; a b$, the scalar components of T at index $i$ effectively scale the inner tensor components of the TT at index $j$. If contracted indices are present, the standard dimensional-matching rules apply.
\item \textbf{Products yielding T (TT $\times$ TT $\to{}$ T)}: It is possible to produce a conventional tensor by performing a full inner contraction for each component-wise multiplication. In operation $i; a b, j; a b \to{} i j$, the inner operation is the tensor dot product $ab,ab \to{}$, which produces a scalar. The result is thus rank-2 T.
\item \textbf{Full contractions (TT dot product)}: Finally, a full dot product between two TTs yields a scalar, as seen in expressions like the PNO-CC energy. The operation $i; a, i; a \to{}$ involves a contraction over the outer index $i$, where each step of the summation involves a tensor dot product ($a,a \to{}$) between the corresponding inner tensors.
\end{itemize}
\end{enumerate}

\subsection{Intermediate Representation (IR) of expressions}\label{sec:IR}

The IR is designed to make several properties of the tensor algebra explicit, which are essential for optimization and numerical evaluation. Specifically, the IR provides a structured format to determine the optimal tensor network contraction ordering, recognize reusable sub-expression evaluations, disambiguate the physical (in-memory) layout of numerical tensor results, and identify the semantics of the underlying tensor operations (see \Cref{sec:operation-basis}).

Structurally, the IR is a full-binary tree, where each node is either a leaf or an internal node with two children. All nodes store data that represents the original expression object along with essential metadata. A key feature of this design is that this metadata can be customized as needed. For instance, when targeting the \texttt{TiledArray} framework, the node data is augmented with tensor indices written as strings suitable for working with \texttt{TiledArray}.

A critical feature of the IR is the mechanism for determining whether two nodes produce an equal numerical result upon evaluation. For a commutative and associative binary operation, many evaluation pathways (binarizations) exist, but all yield the same result. To uniquely identify this result, we define a {\em canonical identity} for each IR node. This identity is established by converting the corresponding IR tree to a {\em canonical form}, which allows for robust identification of equivalent computations even if their binarizations differ. This is an established technique for canonicalizing associative binary evaluation trees, often referred to as associative flattening \cite{aho2007}. However, such techniques do not account for the tensorial structure of the operands, and hence they are insufficient for products of tensors. \code{SeQuant} uses two strategies, one for IR trees that represent tensor networks and another for the rest.

For IR trees that are not tensor networks---such as scalars, products of scalars, or sums of any kind --- the canonical identity is determined by a straightforward recursive procedure. The components of the expression are canonically sorted to handle commutativity and their identities are combined. As scalars have no index structure, their physical layout is trivial and, for a sum of tensors, the resulting index structure is simply inherited from the operands.

In contrast, any IR tree that involves a product of tensors is treated as a tensor network. Even a single tensor is considered a one-node network to properly account for its topology. The canonical identity of such an expression is derived from the canonical representation of its corresponding colored graph (\Cref{sec:canon}). This process yields not only the canonical graph form and an optional phase factor but also a canonical ordering of the network's \emph{external} index slots. This ordering, given by the final vertex order in the canonicalized graph, is a crucial ``byproduct'' of the TN canonicalization that we use to standardize the layout of the tensor/array object representing value of the IR tree. Yet again, symmetries and index dependencies mandate that even for single tensors such layout is determined by full TN canonicalization; naive layout algorithms appropriate for tensors of scalars (bra slots first, then kets) are not sufficient for this purpose. Note that the tensor layout does not necessarily define the actual in-memory layout of the tensor; such details are delegated to the numerical tensor backend.

The power and precision of this graph-based approach are highlighted in \Cref{tab:canon-same-leaf-tot}-\Cref{tab:canon-different-imed-tot}. It correctly identifies algebraically equivalent tensors (\Cref{tab:canon-same-leaf-tot}) and superficially different but equivalent tensor networks (\Cref{tab:canon-same-imed-tot}). Conversely, it can distinguish non-equivalent tensors that happen to share a physical layout (\Cref{tab:canon-different-leaf-tot}), as well as expressions that differ only by a subtle permutation of indices (\Cref{tab:canon-different-imed-tot}).

\begin{table}[!ht]
\setlength{\tabcolsep}{20pt}
\renewcommand{\arraystretch}{1.5}
    \centering
    \caption{Example of two tensors with same canonical identity. The tensors are equivalent modulo a multiplication by a phase and renaming $i_1 \leftrightarrow i_4$.}
    \label{tab:canon-same-leaf-tot}
    \begin{tabular}{lll}
        \toprule
        Tensor & Canonical layout & Phase \\
        \midrule
        ${\tensor*{\bar{g}}{*^{a_2^{{i_3}}}_{i_2}*^{a_3^{{i_4}}}_{a_1^{{i_1}}}}}$ & $i_4 i_3 i_1 i_2 a_1^{i_1} a_3^{i_4} a_2^{i_3}$ & 1 \\
        ${\tensor*{\bar{g}}{*^{a_3^{{i_1}}}_{i_2}*^{a_2^{{i_3}}}_{a_1^{{i_4}}}}}$ & $i_1 i_3 i_4 i_2 a_1^{i_4} a_3^{i_1} a_2^{i_3}$ & -1 \\
        \bottomrule
    \end{tabular}
\end{table}

\begin{table}[!ht]
\setlength{\tabcolsep}{20pt}
\renewcommand{\arraystretch}{1.5}
    \centering
    \caption{Example of two tensors with different canonical identities that produce same canonical layout.}
    \label{tab:canon-different-leaf-tot}
    \begin{tabular}{ll}
        \toprule
        Tensor & Canonical layout \\
        \midrule
        $g_{i_{2}a_{1}^{i_{1}}}^{i_{3}a_{2}^{i_{4}}}$ &
        $i_{1}i_{4}i_{2}i_{3};a_{1}^{i_{1}}a_{2}^{i_{4}}$ \\
        $g_{i_{2}a_{1}^{i_{1}}}^{a_{2}^{i_{4}}i_{3}}$ &
        $i_{1}i_{4}i_{2}i_{3};a_{1}^{i_{1}}a_{2}^{i_{4}}$ \\
        \bottomrule
    \end{tabular}
\end{table}

\begin{table}[!ht]
\setlength{\tabcolsep}{10pt}
\renewcommand{\arraystretch}{1.5}
    \centering
    \caption{Equivalent intermediate tensor expressions and corresponding canonical physical layouts determined by tensor network canonicalization. Indices $i_4$ and $i_5$ have swapped index kinds (batched and external) between the two expressions}
    \label{tab:canon-same-imed-tot}
    \begin{tabular}{ll}
        \toprule
        Intermediate & Canonical layout \\
        \midrule
        $g_{i_{2}i_{3}}^{a_{2}^{i_{4}i_{5}}a_{3}^{i_{4}i_{5}}}s_{i_{4}}^{i_{2}}$
        & $i_{5}i_{3}i_{4};a_{3}^{i_{4}i_{5}}a_{2}^{i_{4}i_{5}}$ \\
        $g_{i_{2}i_{3}}^{a_{2}^{i_{4}i_{5}}a_{3}^{i_{4}i_{5}}}s_{i_{5}}^{i_{2}}$
        & $i_{4}i_{3}i_{5};a_{3}^{i_{4}i_{5}}a_{2}^{i_{4}i_{5}}$ \\
        \bottomrule
    \end{tabular}
\end{table}

\begin{table}[!ht]
\setlength{\tabcolsep}{10pt}
\renewcommand{\arraystretch}{2}
    \centering
    \caption{Non-equivalent intermediate tensor expressions and corresponding canonical physical layouts determined by tensor network canonicalization.}
    \label{tab:canon-different-imed-tot}
    \begin{tabular}{ll}
        \toprule
        Intermediate & Canonical layout \\
        \midrule
        $\left(  g_{i_{2}i_{3}}^{a_{2}^{i_{5}}a_{3}^{i_{4}i_{6}}} s_{i_{5}}^{i_{3}}t_{a_{2}^{i_{5}}}^{i_{5}} s_{i_{6}}^{i_{2}} \right)t_{a_{3}^{i_{4}i_{6}}a_{4}^{i_{4}i_{6}}}^{i_{6}i_{4}}$
        & $i_{6}i_{4};a_{4}^{i_{4}i_{6}}$ \\
        $\left(g_{i_{2}i_{3}}^{a_{2}^{i_{5}}a_{3}^{i_{4}i_{6}}}s_{i_{5}}^{i_{3}}t_{a_{2}^{i_{5}}}^{i_{5}} s_{i_{6}}^{i_{2}} \right) t_{a_{3}^{i_{4}i_{6}}a_{4}^{i_{4}i_{6}}}^{i_{4}i_{6}}$
        & $i_{6}i_{4};a_{4}^{i_{4}i_{6}}$ \\
        \bottomrule
    \end{tabular}
\end{table}

\subsection{IR optimization}

Finding an efficient implementation of tensor algebra expressions is an
optimization problem that involves balancing several competing factors,
including tensor contraction ordering, memory locality, parallelism,
sparsity, and hardware specialization. While a complete solution would
address all these factors, \code{SeQuant}'s contribution is
strategically focused on the high-level symbolic optimizations that are
best performed before an expression is passed to a numerical backend.
This is in contrast to a dedicated tensor algebra compiler, which
operates at a lower level of abstraction. The problem of finding an optimal evaluation strategy for sums of tensor networks that arise in quantum many-body simulation in chemistry has been explored in depth, with several effective strategies emerging. We focus on three distinct classes of optimization: TN contraction ordering (TNCO), TN fusion, and common subexpression elimination (CSE) via intermediate reuse \cite{hartono2005, hartono2006}.

\code{SeQuant}'s approach is tailored to the specific
structure of quantum chemistry problems. It targets contraction ordering
via the proven dynamic programming method, as it is the most effective
strategy for reducing the total operations count and can furthermore use the index extents known at runtime that a general off-line tensor compiler would lack. Furthermore,
the \emph{identification} of common sub-expressions is most effectively
performed at the symbolic level, where \code{SeQuant} can leverage its
TN canonicalizer for identifying common subnetworks.

\code{SeQuant}'s optimization pipeline applies these strategies sequentially. First, TNCO is performed on each individual term of the many-body equations to generate an optimal IR evaluation tree. This approach allows \code{SeQuant} to apply the most effective optimization (TNCO) to each term individually while still capturing global CSE opportunities through its robust tensor network canonicalizer during the subsequent interpretation phase. Future work might explore coupling these two optimization stages for global optimization, as was done by some \cite{VRG:engels-putzka:2011:JCP}, but such optimizations are unlikely to be viable at runtime and would need to be performed offline. The remaining factors—parallelism, sparsity, data locality, and hardware specialization—are delegated to a high-performance numerical backend, which in our case is \code{TiledArray}. This strategic division of labor allows \code{SeQuant} to produce an efficient evaluation plan for the numerical library to execute.

\subsubsection{Tensor network contraction ordering (TNCO)}

Finding the optimal sequence of binary contractions for a tensor network, known as the TN contraction ordering (TNCO) problem\footnote{Although the tensor network contraction ordering problem was originally considered in the context of tensor algebras with covariant tensor networks only, thus all products being contractions (or outer products), here we apply it more broadly to tensor networks involving general tensor products.} problem, is NP-complete \cite{chi-chung1997}. However, the tensor networks encountered in quantum chemistry are typically composed of a relatively small number of tensors (often fewer than 10). Consequently, an exhaustive search for the optimal contraction order using dynamic programming is computationally feasible and highly effective \cite{VRG:lai:2012:CPL}. Other works have focused on TNCO for single, large tensor networks prevalent in quantum many-body physics \cite{VRG:pfeifer:2014:PRE}, with recent efforts exploring the use of machine learning to solve the TNCO problem \cite{meirom2022,xu2024}.

\code{SeQuant} implements TNCO using a bottom-up dynamic programming
algorithm\cite{hartono2005, hartono2006, VRG:lai:2012:CPL}, which can be
 conceptually understood as a recursive process
for finding the optimal binarization of a tensor network.
For networks with three or more tensors, the algorithm considers all
unique, non-trivial bipartitions of the set of tensors. A non-trivial
bipartition divides the network into two non-empty tensor sub-networks.
For example, a three-tensor network $A B C$ has three such
bipartitions: $(A, BC)$, $(B, A C)$, and $(A B, C)$. For each
bipartition, the total operation count is calculated as the sum of
three costs: the cost of contracting the two intermediate tensors
resulting from each partition, plus the recursively determined optimal
costs for evaluating the partitions themselves. The final optimal
binarization for the original network is then the one constructed from
the bipartition that yields the minimum total operations count.

\subsubsection{Tensor Network Fusion}

Another class of symbolic optimization, which we term \emph{tensor network fusion}, transforms a sum of terms by factoring out a common tensor network. For example, the expression:
$\frac{1}{16}(\bar{g}_{k l}^{c d} t_{c d}^{i j}) t_{a b}^{k l} + \frac{1}{8}(\bar{g}_{k l}^{c d} t_{c d}^{i j}) t_{a}^{k} t_{b}^{l}$
can be fused into a more compact form:
$\frac{1}{16}(\bar{g}_{k l}^{c d} t_{c d}^{i j})(t_{a b}^{k l} + 2 t_{a}^{k} t_{b}^{l})$.
This transformation reduces the total number of expensive tensor contractions. In general-purpose compilers, such re-associations are often considered ``unsafe" optimizations due to potential floating-point inaccuracies and are typically disabled by default. However, for the specific domain of quantum chemistry, this strategy has been explored \cite{hartono2006, VRG:lai:2012:CPL, VRG:engels-putzka:2011:JCP}.

Finding the optimal fusion scheme is a difficult combinatorial problem, especially when coupled with TNCO. The number of tensor networks to be considered for fusion can range from tens to thousands, making an exhaustive search impractical. Furthermore, the potential performance gains are often less dramatic than the orders-of-magnitude savings from optimal TNCO. For these reasons, an automated tensor network fusion algorithm has not been prioritized in the current implementation of \code{SeQuant}.

Nevertheless, the framework does support fusion of two tensor networks that share a common tensor sub-network, allowing for targeted, user-driven optimizing rewrites. Future work may explore heuristic-based strategies \cite{hartono2006, VRG:engels-putzka:2011:JCP} to achieve a more widely applicable automation of this optimization.

\subsubsection{Common sub-expression elimination (CSE)}

Given the complexities of fusion, \code{SeQuant} prioritizes a more general optimization: the identification and reuse of common intermediates at evaluation time. Although this reduces the computational cost by only a factor, the practical savings in resources—time, memory, and communication—are often substantial. The critical identification of common sub-expressions is performed at the symbolic level, while the reuse occurs at runtime.

This approach is best understood as a multi-step evaluation plan. For
the same example expression:
\begin{enumerate}
  \item Identify and evaluate the common intermediate once:
	    $I_{k l}^{i j} \leftarrow \bar{g}_{k l}^{c d} t_{c d}^{i j}$.
  \item Reuse the intermediate in all subsequent operations:
  \begin{itemize}
    \item Evaluate $\frac{1}{16} I_{k l}^{i j} t_{a b}^{k l}$.
    \item Evaluate $\frac{1}{8} I_{k l}^{i j} t_{a}^{k} t_{b}^{l}$.
  \end{itemize}
\end{enumerate}
Intermediates are stored in a cache that also help control the intermediate lifetime by tracking each intermediate's pending use counter; once it reaches zero intermediate is removed from the cache.
This strategy is in some sense more general than fusion because it can identify common sub-expressions across \emph{different equations} and \emph{within single tensor networks}. \code{SeQuant}'s TN canonicalizer (\cref{sec:canon}) solves the primary challenge: robustly identifying computationally identical tensor (sub)networks despite arbitrary dummy index labeling or ordering. By computing a unique canonical representation for each tensor network, multiple instances can be detected before the optimized evaluation plan is delegated to the numerical backend.

\subsection{Runtime Interpretation}\label{sec:interpretation}

The numerical evaluation of the optimized expression is performed by a tree-walk interpreter that operates directly on the IR. This process faces two main challenges. First, the evaluation results are heterogeneous; IR nodes can represent different types, from simple scalars to complex framework-specific tensor structures. Second, the implementation of tensor operations, such as contraction, varies significantly between different numerical tensor libraries.

To address these challenges, we introduce a \code{Result} interface. This class serves as a type-erased wrapper for the underlying numerical data, simplifying the interpreter's design by providing a common interface for the fundamental operations corresponding to the IR nodes (e.g., sum, product, permutation).

The specific implementation of these operations is delegated to derived classes tailored to each supported numerical framework. This approach ensures that the system is extensible to new backend libraries. For instance, to demonstrate this flexibility, a backend for the \code{BTAS} library\cite{BTAS1} has been implemented for expressions involving tensors of scalars, complementing the primary \code{TiledArray} framework. Adding additional backend by analogy with the existing backends is completely formulaic and can be efficiently accomplished by modern coding agents; starting from the \code{BTAS} backend code Anthropic's Claude 4.6 coding agent developed evaluation backend based on the new BLAS-like TAPP API\cite{brandejs2026tensoralgebraprocessingprimitives} in less than 30 minutes.

The interpretation, shown in pseudocode in \Cref{algo:interpret}, is a recursive post-order traversal of the IR tree. The process at any given internal node begins by querying a \emph{cache manager} using the node's \emph{canonical identity} as a key. If a result exists in the cache, it is returned immediately; this serves as the runtime implementation of the CSE optimization. If the result is not cached, the interpreter recursively calls itself on the node's children. The recursion ends at the leaf nodes. At a leaf, a \emph{leaf evaluator} function is invoked to retrieve the raw numerical data. This data is then permuted into the canonical memory layout that was determined by the graph canonicalization process described in the previous section (see \Cref{sec:canon}). Once the recursive calls return, the results from the child nodes are available in their respective canonical layouts. The parent node then performs its specified operation (e.g., sum or product). The layout of the newly created intermediate tensor is also determined by its pre-computed canonical form. This new result is stored in the cache using the parent's canonical identity as the key before being returned. All underlying numerical operations are delegated to a backend framework.

This interpreter-based approach using the \code{TiledArray} backend has been validated in production within the \code{MPQC}\cite{VRG:peng:2020:JCP} quantum chemistry package, where it provides a robust and efficient pathway for the automated method implementation. Detailed assessment of the performance of the interpreted tensor expressions will be presented in the follow-up manuscript.

\begin{algorithm}
    \centering
  \includegraphics[width=\linewidth]{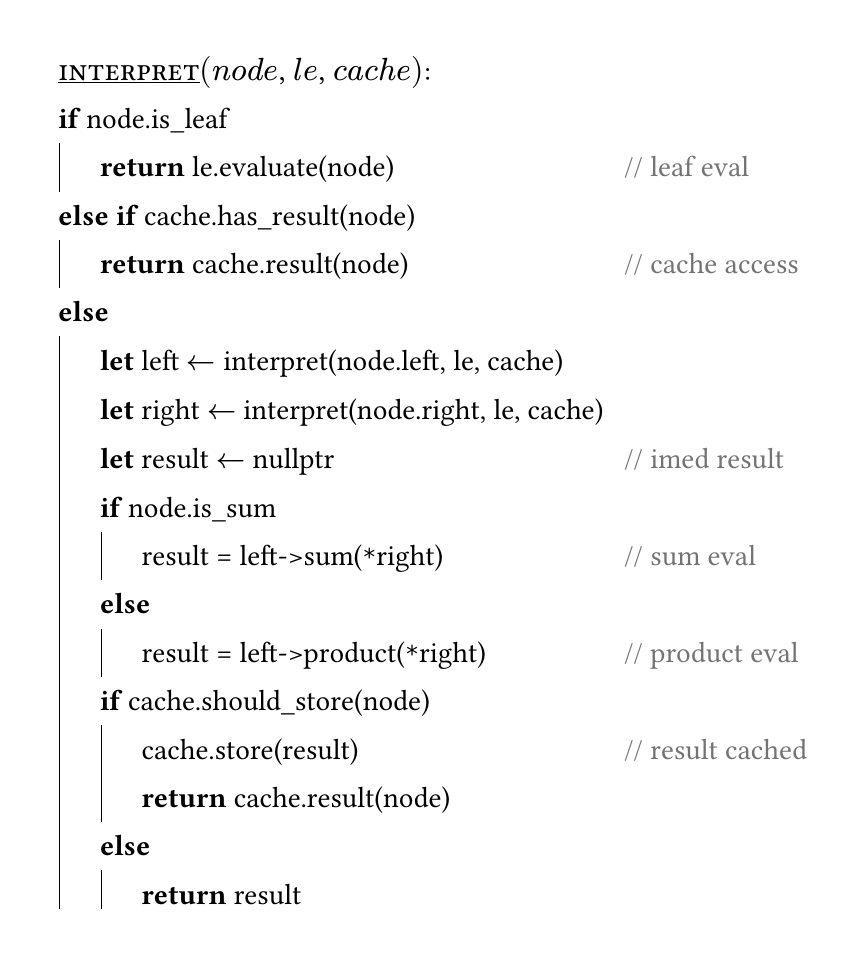}
    \caption{Overview of the tree-walk interpretation with cache management}
    \label{algo:interpret}
\end{algorithm}

\section{Summary}\label{sec:summary}

\code{SeQuant} is an open-source library for symbolic algebra of tensors over commutative (scalar) and non-commutative (operator) rings.
It is implemented in C++ to maximize its portability, performance, and reusability as a runtime component wherever symbolic tensor algebra is needed.
The \code{SeQuant} source code is available at \url{https://github.com/ValeevGroup/SeQuant}.

This manuscript described the core features of \code{SeQuant} applicable to any domain. The key innovation supporting most of its functionality is a graph-theoretic tensor
network (TN) canonicalizer. The use of graph-theoretic approaches for tensor canonicalization is not novel, \cite{VRG:obeid:2001:,VRG:bolotin:2013:,VRG:li:2017:P2AISSAC,VRG:peeters:2018:J,VRG:kryukov:2019:JPCS} 
the ability to support noncovariant tensor networks and nested index dependencies (\cref{sec:tensor:index_dependencies}) is novel. The former occur, for example, in expressions involving summation over an index with 3 or more slots that arise from tensor factorizations; such indices represent hyperedges in the tensor network.
The latter occurs in, e.g., block-wise data compressions in data science and modern quantum simulation. The TN canonicalizer is used throughout \code{SeQuant} for routine simplification of conventional (scalar) tensor expressions. It is also used to optimize the application of Wick’s theorem, which canonicalizes products of tensors over operator
fields. \code{SeQuant} blurs the line between Wick's theorem application and the diagrammatic techniques whose success is closely related to their graph-theoretic roots; the use of topological information about equivalence of the index slots and tensors in a tensor network produced by the graph-theoretic canonicalizer allows to accelerate the algebraic Wick's theorem engine to reach the speed of diagrammatic techniques without any context specific optimizations. 

\code{SeQuant} blurs yet another
line between pure symbolic manipulation/code generation and numerical evaluation by including compiler-like components to optimize and directly interpret tensor expressions using external numerical tensor
algebra frameworks. The routine use of interpretation, rather than code generation, allows us to greatly speed up the development cycle by avoiding the need to recompile the code after each change. Runtime interpretation also allows us to exploit the available runtime context (such as the extents of index spaces for the particular computation) when optimizing the expressions for evaluation. Key elements of symbolic manipulation of the expressions in preparation for their numerical evaluation also exploit the efficient tensor canonicalizer, for computing the canonical identity of subexpressions (essential, for example, for common subexpression elimination) and for determining the canonical layout of tensors in memory.

Additional features of \code{SeQuant} used for quantum many-body simulation as well as representative performance of \code{SeQuant}-driven implementations thereof will be described in the second manuscript in the series.
Such features, together with the code functionality described here, significantly accelerate the development cycle of scientific applications that require heavy-duty symbolic manipulation and numerical evaluation of tensor algebra, from the formal derivation of a new theory to its correct high-performance numerical implementation.

\begin{acknowledgments}
This research was supported by the US National Science Foundation via Award 2217081.
The work by EFV was supported by the US Department of Energy, Office of Science, via award DE-SC0022327. The work by RGA was partially supported by a Kekulé fellowship of the Fonds der Chemischen Industrie.
\end{acknowledgments}

\section*{Data Availability Statement}

The source code of \code{SeQuant} is available at \url{https://github.com/ValeevGroup/SeQuant}. Version 2.2.0 of \code{SeQuant} described in this article is available from Zenodo via DOI 10.5281/zenodo.18689273\cite{valeev_2026_18689273}. Raw data from the computational experiments are available from the authors upon reasonable request.



\bibliography{vrgrefs,references}

\end{document}